%% file: Main.tex
\documentclass[acmsmall,screen]{acmart}
\AtBeginDocument{%
  \providecommand\BibTeX{{%
    \normalfont B\kern-0.5em{\scshape i\kern-0.25em b}\kern-0.8em\TeX}}}

\setcopyright{acmcopyright}
\copyrightyear{2024}
\acmYear{2024}
\acmDOI{}

\acmConference[MobileHCI '24]{MobileHCI '24: ACM MobileHCI}{Sept 30 -- Oct 03, 2024}{Melbourne, Australia}
\acmBooktitle{Proceedings of the ACM on Human-Computer Interaction: ACM MobileHCI,
  Sept 30 -- Oct 03, 2024, Melbourne, Australia}
\acmISBN{}

\usepackage{array}
\usepackage[section]{placeins}
\usepackage{subcaption}
\usepackage{graphicx}
\usepackage{wrapfig}

\begin{document}

\title{FacePsy: An Open-Source Affective Mobile Sensing System -- Analyzing Facial Behavior and Head Gesture for Depression Detection in Naturalistic Settings
}


\author{Rahul Islam}
\email{mislam5@stevens.edu}
\orcid{0000-0003-3601-0078}
\author{Sang Won Bae}
\authornote{Corresponding author.}
\orcid{0000-0002-2047-1358}
\email{sbae4@stevens.edu}
\affiliation{%
  \institution{Charles V. Schaefer, Jr. School of Engineering and Science, Stevens Institute of Technology}
  \streetaddress{1 Castle Point Terrace}
  \city{Hoboken}
  \state{New Jersey}
  \country{USA}
  \postcode{07030}
}

\renewcommand{\shortauthors}{Islam and Bae}

\begin{abstract}
Depression, a prevalent and complex mental health issue affecting  millions worldwide, presents significant challenges for detection and monitoring. While facial expressions have shown promise in laboratory settings for identifying depression, their potential in real-world applications remains largely unexplored due to the difficulties in developing efficient mobile systems. In this study, we aim to introduce FacePsy, an open-source mobile sensing system designed to capture affective inferences by analyzing sophisticated features and generating real-time data on facial behavior landmarks, eye movements, and head gestures -- all within the naturalistic context of smartphone usage with 25 participants. Through rigorous development, testing, and optimization, we identified eye-open states, head gestures, smile expressions, and specific Action Units (2, 6, 7, 12, 15, and 17) as significant indicators of depressive episodes (AUROC=81\%). Our regression model predicting PHQ-9 scores achieved moderate accuracy, with a Mean Absolute Error of 3.08. Our findings offer valuable insights and implications for enhancing deployable and usable mobile affective sensing systems, ultimately improving mental health monitoring, prediction, and just-in-time adaptive interventions for researchers and developers in healthcare.

\end{abstract}


\keywords{Affective computing, Depression, Machine Learning, Mobile computing, System, Empirical study that tells us about people, Application Instrumentation, Field Study}



\maketitle

\input{Section/1_Introduction}

\input{Section/2_Related_Work}

\input{Section/3_Method}

\input{Section/4_Results}

\input{Section/6_Discussion}
\input{Section/7_Limitation}

\input{Section/9_Conclusion}

\section{Acknowledgments}
We sincerely thank our research volunteer, Shahnaj Laila, for her kind assistance in conducting the feasibility study. We are also grateful to the
participants who generously agreed to share their facial behavior data for
this study.

\bibliographystyle{ACM-Reference-Format}
\bibliography{Reference}

\input{Section/10_Appendix}

\end{document}

%% file: Section/1_Introduction.tex
\section{Introduction}
Mental health pertains to emotional, psychological, and social well-being, influencing daily thoughts, feelings, and actions. Mental illness is a leading cause of disability, with an estimated 450 million people affected worldwide \cite{world2003investing}. It is also a significant predictor of suicide \cite{nock2010mental}. Mental disorders usually emerge in an individual's early 20s \cite{kessler2005lifetime}, and their untreated presence can negatively impact academic success, productivity, and social relationships \cite{kessler1995social, wang2007telephone}. In the context of COVID-19, the need for social distancing has led to the widespread adoption of telehealth services such as telepsychiatry \cite{telehealthscoping, o’brien_mcnicholas_telepsych}. Unfortunately, many individuals are experiencing mental health issues during the pandemic, with the highest levels of pandemic-era anxiety and depression observed in 2020 across all age groups, which began to decline in early 2021 \cite{villaume2023age}. While many COVID-19 restrictions have been lifted, approximately 5\% of the U.S. adult population, or about 12 million Americans, are living with co-occurring chronic pain and clinically significant symptoms of anxiety and depression \cite{jennifer2024co}. This ongoing situation underscores the importance of reconsidering how to deliver mental health care effectively at the right time. Personalized psychiatric care, which promotes preventive measures and offers tailored interventions, could help meet these needs, though its availability remains limited \cite{abdullah2018sensing}.

A growing body of psychological studies \cite{ellgring2007non, girard2013social, girard2014nonverbal, stuhrmann2011facial} have suggested that depression is characterized by nonverbal signals such as facial muscle movement, and head gesture, which can be detected automatically without the need for clinical intervention. We refer to these as \textit{"facial behavior primitives"}. Research has shown that mental illness, such as depression, leaves recognizable markers in the facial patterns of an individual \cite{samareh2018detect}. Often, these changes manifest in a person's face involuntarily. Creating an automatic system \cite{cohn2009detecting, song_spectral_2020, valstar2014avec} based on these cues can provide an objective and repeatable evaluation and address problems related to cost and time requirements. Despite the valuable insights gained from these studies, it should be noted that they were conducted in controlled lab environments and recorded videos of individuals' faces. Currently, the real-life implementation of such systems is limited due to privacy concerns \cite{cohn2009detecting, valstar2013avec, song2020spectral, kong2022automatic, casado2023depression}, unrealistic costs \cite{pedrelli2020monitoring}, and required computational power \cite{chikersal2021detecting, ferreira2015aware, opoku2022mood}.

While facial actions have shown promise in lab settings for understanding depression, their application in real-world scenarios remains largely unexplored due to challenges in designing efficient, deployable mobile systems. To bridge this gap, we introduce FacePsy, an open-source mobile sensing system capturing facial features, generating real-time data on facial behavior landmarks, eye open, smile, and head gestures, all through smartphones in natural settings, while preserving user privacy. We hypothesize that digital biomarkers extracted from facial cues offer valuable insights into an individual's internal emotional and affective state, thereby enabling algorithms to infer depression. In this field study, we gathered data in real-world contexts to assess how and whether the data collected through our framework could demonstrate the potential for predicting depressive episodes in naturalistic environments.

While there are studies that use mobile sensing to track patterns in social and behavioral data by tracking communications, app usage, and GPS data \cite{wang2014studentlife, chikersal2021detecting, opoku2022mood}, these mobile sensing-based solutions primarily focus on capturing social and behavioral data but disregarding affective signals, which have been shown to be important indicators of depression. These studies have particular challenges and barriers: (1) Mobile sensing limits modeling usage because it requires extensive computation and post-processing. The burden of collecting sensors 24/7 and battery consumption may lead to low compliance. It might not be usable for stakeholders. (2) Wang et al. \cite{wang2015using} tried to capture entire face images in the real-world settings to understand depression but reported that they failed to validate the effectiveness of data due to insufficient frames to build a model (one frame when unlocking the smartphone ); (3) Tseng et al. collected and analyzed eye patches in detecting alertness \cite{tseng_alertnessscanner_2018} not depression, but partially captured a part of the face only (eye). Most recently, MoodCapture \cite{nepal2024moodcapture} was introduced that captures facial images in natural environments for depression detection. This involves analyzing image attributes such as angle, dominant colors, location, objects, and lighting. The utility of MoodCapture for developers seeking to implement similar studies in different settings may be somewhat limited, as the authors have not made their mobile system, dataset, or machine learning pipeline publicly available. Our study complements MoodCapture's work by exploring the incremental utility of mobile sensing for depression detection and advocating for new avenues to develop mental health assessment tools based on in-the-wild images. Our research advances from MoodCapture in terms of data collection mechanisms, on-device processing, privacy awareness, and the facial attributes collected. While MoodCapture collects facial data when a user responds to survey questions, our study implements a trigger-based data collection motivated by prior literature \cite{tseng2018alertnessscanner}. This mechanism activates based on user actions, such as turning the screen on/off, opening/closing apps, etc., to start or stop data collection. For on-device processing, we only send the final detected facial behavior primitives (Action Units (AU), smile, eye open state, head Euler angles, and landmarks) to the research server for further analysis. This helps us ensure user privacy and prevent the leakage of facial images by discarding images from user device after processing. Our work also advocates for privacy-aware data collection. This approach is informed by prior literature on nudging \cite{balebako2014improving}, informing users about active data collection \cite{felt2012ve, denning2014situ}, and notifying users when the app restarts itself after a reboot or crash \cite{denning2014situ}. While MoodCapture has first introduced the use of facial images for depression detection in natural environments, our approach proposes a novel facet to this field in several key aspects. We are among the first to develop an open-sourced, privacy-aware, trigger-based affective mobile system that captures facial data from users' smartphones and immediately discards the raw images after extracting essential features in near real-time (within 10 seconds). This method not only builds a predictive model of depressive episodes but also ensures that sensitive facial data is not permanently stored on devices, addressing significant privacy concerns and awareness.

To advance affective computing systems, making it applicable in real-world settings, we synthesize a set of affective signals from face which have been well-validated such as facial muscular activities (AUs) as well as proposed new features beyond simple facial expression algorithms that have been unexplored and invalidated in a user's everyday settings. Research questions as follows: (RQ1) What are the important signals of affective biomarkers on depressive episode detection by differentiating depressive and non-depressive episodes, and how can those key features contribute to the model's performance? and (RQ2) Whether and how can an affective mobile system be efficiently designed, tested, and developed to understand a user's mental health, specifically in predicting depressive episodes, in real-world settings? As a result, we identified specific affective indicators as crucial factors for distinguishing between individuals experiencing depressive and non-depressive episodes. These indicators encompass the eye-open state, head pose, smile expression, and specific Action Units (2, 6, 7, 12, 15, and 17). When these features are combined, they exhibit predictive potential for detecting depression episodes, achieving an AUROC of 67\% for universal model, while the hybrid model has an AUROC of 81\%. It is worth noting that further enhancements in predictive accuracy can be attained through the accumulation of additional data spanning subsequent weeks. These findings represent a significant stride in bridging the existing disparity between controlled laboratory studies and the practical implementation of depression detection through affective mobile sensing systems in real-world scenarios.

As such, we have developed a deployable and usable open-sourced, lightweight, affective mobile system for the HCI community \footnote{{Our system source code is available at: https://github.com/stevenshci/FacePsy}}. Our system integrates state-of-the-art facial biomarkers, which have been validated to understand complex mental states and workloads in controlled lab settings. We have further expanded these features for the context of depression detection. The system automatically extracts these features from a user's smartphone in natural environments. This system has the potential to create new avenues for developing mental health assessment tools and behavior modeling based on in-the-wild images. Moreover, our FacePsy system can be deployed in everyday settings, and it is optimized with a sampling rate of 2.5 FPS without any delays on the user's own phones. Further, we provide insights with the experiments with different subset of facial behavior primitives features in detecting depression in naturalistic environments. As we highlight the impact of each subset of features validated in naturalistic environments, researchers and developers can utilize our mobile system to conduct their studies, configure apps for triggering time and frequency, and build their own computational models.

%% file: Section/2_Related_Work.tex
\section{Background}
In this section, we introduce the literature on facial behavior primitives in developing depression inferences and machine learning models (Section \ref{FacialBehaviorPrimitives}) and provide prior work on depression detection using mobile sensing technologies in the fields of mobile and affective computing communities (Section \ref{DepressionMobileSensing}).

\subsection{Facial Behavior Primitives in Depression } \label{FacialBehaviorPrimitives}
Research on nonverbal facial behavior often shows that individuals with depression usually exhibit fewer happy facial expressions, reduced expressiveness, and less head movement. The less frequent display of happy facial expressions by depressed patients is a commonly observed finding \cite{chentsova2010further, gehricke2000reduced, renneberg2005facial, sloan2001diminished}. Various studies also link depression with decreased general facial expression \cite{renneberg2005facial, gaebel2004facial} and head movement \cite{fisch1983analyzing, joshi2013can, alghowinem_head_2013}. One study found that participants with major depressive disorder (MDD) had a significantly reduced transient pupillary response \cite{laurenzo2016pupillary}, lending support to the potential of detecting depression through these means. Typical symptoms of depression, such as sorrowful expressions and a lack of affective experience, have been characterized by researchers using facial expressions \cite{ellgring2007non}. However, the prevalence of negative facial expressions in depressed individuals is disputed, with contradictory findings presented in various studies. Some argue that depression is characterized by an increase in negative facial expressions \cite{sloan1997subjective, reed2007impact}, while others suggest that depressed people may actually exhibit more positive facial expressions \cite{renneberg2005facial, gaebel2004facial}.

To diagnose depression, nonverbal signals have been introduced by researchers in affective computing. For instance, Cohn et al. \cite{cohn2009detecting} proposed visual signals as non-verbal behavioral features -- manually annotated facial action units (AUs) and active appearance model (AAM) features, which are mathematically derived representations of facial images that capture shape and texture variations. They found that participants with high depression severity displayed fewer associative facial expressions (AU12 -- lip corner puller, and AU15 -- lip corner depressor) and more non-associative facial expressions (AU14 -- dimpler), indicating that these traits helped spot depression. Most recently, Valstar et al. \cite{song2020spectral, valstar2014avec} have used facial and auditory features to detect depression in pre-recorded videos. Their finding suggests that AU4 (brow lowerer), AU12 (lip corner puller), AU15 (lip corner depressor), and AU17 (chin raiser) are useful for estimating depression severity, supporting existing evidence. These findings highlight the potential utility of automated facial behavior analysis towards predicting of depression. These initial studies were based on recorded video data from consenting participants in controlled environments. In contrast, deploying such technologies in uncontrolled, everyday settings raises valid privacy concerns. However, our approach mitigates these concerns by processing data directly on the device. By leveraging on-device computation for feature extraction, we significantly reduce the privacy risks associated with transmitting sensitive facial data. This method ensures that personal data does not leave the user's device, aligning with privacy-preserving strategies essential for real-world applications.

\begin{table}[h]
\caption{\label{tab:background}Studies on Predicting Depression Using Facial Behavior Primitives}
\centering  
\tiny

\begin{tabular}{p{1.5cm}p{.5cm}p{1.5cm}p{3cm}p{1.5cm}p{2cm}p{1.3cm}}
\toprule
\textbf{Study} & \textbf{Part.} & \textbf{Study Length} & \textbf{Data \& Feature Types} & \textbf{Validation Method} & \textbf{Performance Metrics} & \textbf{Research Environment} \\
\midrule

Cohn et al. \cite{cohn2009detecting} & 57 & 7-week intervals & Audio/video, 17 AUs, AAM & Leave-one-out & Accuracy: 79\% & Lab \\

Valstar et al. \cite{valstar2013avec} & 292 & One video each & AVEC13, LPQ & 5-fold & MAE: 10.88, \newline RMSE: 13.61 & Lab \\

Song et al. \cite{song2020spectral} & 84 & One video each & AVEC14, 17 AUs, Pose, Gaze & 50/50 split & MAE: 5.95, \newline RMSE: 7.15 & Lab \\

Kong et al. \cite{kong2022automatic} & 102 & N/A & 10 photos, Deep learned & 7:2:1 split & Accuracy: 98.23\% & Lab \\

Casado et al. \cite{casado2023depression} & 376 & One video each & AVEC13/14, rPPG & N/A & AVEC13: MAE: 6.43, \newline AVEC14: MAE: 6.57 & Lab \\

Wang et al. \cite{wang2015using} & 37 & 10 weeks & Photos, Eigenfaces, landmarks & N/A & N/A & In the wild \\

Nepal et al. \cite{nepal2024moodcapture} & 177 & 90 days & AU, Gaze, Head Pose, Rigidity Parameters, and Eye, 2D \& 3D Landmarks & 5-fold leave-subject-out & Balanced Acc: 61\% & In the wild \\

Our approach & 25 & 4 weeks & 12 AUs, Smile, Eye open, Head Pose, EAR, IVA, 133 landmarks & Leave-One-Person-out & Accuracy: 51\%, \newline AUC: 67\% \newline MAE: 3.26 & In the wild \\
 &  &  &  & Leave-One-Day-out & Accuracy: 69\%, \newline AUC: 81\% \newline MAE: 3.08 &  \\
\bottomrule
\end{tabular}
\end{table}

Researchers in HCI have explored using mobile camera sensors to capture and analyze user data for mental health assessment. For instance, Rui et al. \cite{wang2015using} developed a smartphone app that takes photos of users' faces throughout the day, extracting facial expressions and landmarks. However, this approach had limited success. The correlation between facial expressions, landmarks, and mental health was difficult to establish due to the poor performance of facial expression algorithms, landmark detectors, and image quality. In a separate HCI study, Vincent et al. \cite{tseng2018alertnessscanner} used pupil information to gauge user alertness as an indicator of mental states. A recent study, MoodCapture \cite{nepal2024moodcapture}, evaluated depression using images taken automatically by smartphone front-facing cameras during everyday activities. This tool analyzes features in the images such as angles, dominant colors, locations, objects present, and lighting conditions. The study showed that a random forest algorithm trained on facial landmarks can effectively distinguish between depressed and non-depressed individuals, and predict raw PHQ-8 scores. The effectiveness of MoodCapture for developers looking to conduct similar studies in various environments might be constrained because the authors haven't released their mobile system, dataset, or machine learning pipeline. Our study advances from MoodCapture in terms of data collection mechanisms, on-device processing, privacy awareness, and the facial attributes collected. While MoodCapture captures images when participants respond to EMA questions, our protocol relies on opportunistic data collection. We gather data when participants interact with their smartphones at specific triggers (See Section \ref{sec:UBS}). This allows us to collect information that more comprehensively represents participants' daily lives and emotional states. By using a broader data collection framework, we can conduct a richer analysis of behavioral patterns and emotional nuances that occur during regular phone usage, not just during specific survey responses. Both MoodCapture and our study are primarily designed for depression detection. However, using smartphone cameras to capture images could also extend to assessing other cognitive states, such as alertness, through pupil imaging. These approaches \cite{tseng2018alertnessscanner, wang2015using, nepal2024moodcapture}, however, raises privacy concerns due to the transmission and processing of facial images on external servers. Our study complements the findings of these studies by implementing all data processing locally on the user's device; we plan to open source our system to the research community, which can be used for further studies in behavior modeling through affective signals.

\subsection{Detecting Depression Using Mobile Sensing} \label{DepressionMobileSensing}
Significant progress has been made in detecting depression with mobile sensing. For instance, Chikersal et al. \cite{chikersal2021detecting} utilized the AWARE \cite{ferreira2015aware}, an open-source context instrumentation framework. It tracked behavioral data such as Bluetooth, calls, GPS, microphone, and screen status from smartphones and wearable fitness devices to detect depression in college students. Their method achieved 85.4\% accuracy in identifying changes in individuals' depression and 85.7\% accuracy in detecting post-semester depression over a semester-long (16 weeks) study. In a different study, Asare et al. \cite{opoku2022mood} also used the AWARE framework to monitor behavioral data. They focused on sleep, physical activity, phone usage, GPS location, and daily mood ratings using the circumplex model of affect (CMA) to detect depression. This approach resulted in 81.43\% accuracy. Though sensing systems have proven effective in various applications, their ability to provide real-time insights and interventions is relatively unexplored. This is mainly due to the resource-intensive requirements of pre-processing, feature engineering, and model development, which demand further scrutiny to fully utilize their capabilities. It is worth noting that the systems examined in previous research did not support near-real-time facial feature extraction. This distinctive feature separates our affective mobile system from others. We summarize the findings of these studies in Table \ref{tab:background2}.

\begin{table}[h]
\caption{\label{tab:background2}Studies on Predicting Depression Using Mobile Sensing}
\centering
\tiny

\begin{tabular}{p{2cm}p{3cm}p{2cm}p{5.5cm}}
\toprule
\textbf{Study} & \textbf{Sensors Used} & \textbf{Results} & \textbf{Findings} \\
\midrule

Chikersal et al. \cite{chikersal2021detecting} & Bluetooth, GPS, Screen, Calls, Sleep & Accuracy: 82.3\% \newline F1: 78\% & Identifies depressive symptoms using data from smartphones and fitness trackers of college students. The study introduce advanced feature extraction technique. \\

Opoku et al. \cite{opoku2022mood} & Sleep, Activity, GPS, Phone Usage & Accuracy: 81.43\% \newline AUC: 82.31\% & Classifies individuals as depressed/non-depressed using mood scores and sensor data. Significant differences found in mood, sleep, activity, phone usage, GPS mobility. \\

Pedrelli et al. \cite{pedrelli2020monitoring} & EDA, Heart Rate, Accelerometer, Sleep, Movements, Temperature & MAE: 3.88 - 4.74  & Feasibility of monitoring depression severity with smartphones and wearables, showing moderate to high correlations with clinician-assessed scores. \\

Farhan et al. \cite{farhan2016behavior} & GPS, Physical Activity & F1: 55\% & Behavioral data from smartphones predict clinical depression. Combining with PHQ-9 scores enhances accuracy. \\

Nepal et al. \cite{nepal2024moodcapture} & AU, Gaze, Head Pose, Rigidity Parameters, Eye, 2D \& 3D Landmarks & Balanced Acc: 61\% & Uses smartphone images to detect depression by analyzing facial expressions and features, demonstrating machine learning potential in mental health assessment. \\

Islam and Bae \cite{islam2024pupilsense} & Pupil-Iris Ratio & Accuracy: 76\% \newline F1: 64\% \newline AUC: 71\% & Pupillary response in natural settings varies between morning and evening and can differentiate between depressive and non-depressive states. \\

Our approach & 12 AUs, Smile, Eye open, Head Pose, EAR, IVA, 133 landmarks & Accuracy: 69\% \newline F1: 67\% \newline AUC: 81\% \newline MAE: 3.08 & FacePsy detects depressive episodes by collecting facial behaviors and head gestures in real-world settings, achieving high predictive accuracy with key features like eye-open states, smile expressions, and specific Action Units. \\

\bottomrule
\end{tabular}
\end{table}

Pedrelli et al. \cite{pedrelli2020monitoring} have collected data streams from a wearable tracker, Empatica, and smartphone to detect changes in depression severity. The authors leveraged physiological data such as electrodermal activity (EDA), peripheral skin temperature, heart rate, motion from the 3-axis accelerometer, sleep characteristics, social interactions, activity patterns, and the number of apps used, etc. They evaluated their predictive models using two evaluation methods: user-split and time-split. They achieved an mean absolute error (MAE) ranging between 3.88 and 4.74. However, their work's limitation is that participants must wear two E4 Empatica, one in each hand. Such obtrusive approaches lead to decreased compliance of participants and extra cost (\$1,690 per device) for researchers and users. In another study \cite{farhan2016behavior}, they have used GPS and physical activity as sensor features for depression. In another study \cite{islam2024pupilsense}, researchers used pupillometry data as a proxy for psychological state to detect depression. They found pupillary response in natural settings varies between morning and evening and can differentiate between depressive and non-depressive states. All of the earlier studies have mostly focused on behavioral and social changes in a person during the depression because open-sourced and deployable mobile frameworks are limited in the HCI community. As we know, depression is a multifaceted disorder that affects the behavioral, physiological, and social aspects of people's lives. The current mobile sensing-based approaches \cite{opoku2022mood, chikersal2021detecting, farhan2016behavior} may not be able to capture the physiological signals with rich emotional signals, which have been shown to be important indicators of depression. Whereas wearable sensing-based physiological sensing solutions are proposed, the cost is very high for such deployment. As motivated by the works \cite{cohn2009detecting, valstar2013avec, song2020spectral, kong2022automatic, casado2023depression} in affective computing have shown great potential in capturing physiological signals with rich emotional data in persons with depression in lab settings. However, these studies have been unexplored in real-world deployment where continuous symptom monitoring is essential part of delivering an appropriate real-time intervention. In this paper, we would address this specific issue by developing an affective mobile system that can unobtrusively and opportunistically track individual facial behavior primitives to give insights into their complex mental state in near-real time by extracting various emotional data motivated by theories in affective computing. As human face serves as a crucial and natural medium for conveying emotional and mental states \cite{el2005real}. While some studies use facial behavior primitives to detect depression \cite{song2020spectral, valstar2014avec}, attempts to detect depression using passively sensed facial behavior primitives in a naturalistic environment have been unsuccessful \cite{wang2015using}. The effectiveness of the existing depression detection models based on facial behavior primitives in natural environments is also unexplored.

Therefore, we aim to design, refine, and develop configurable triggering for data collection, including when unlocking phones and app use, to capture users' facial behavior data in their everyday settings. Our method opportunistically captures users' facial behavior primitives unobtrusively by triggering data collection when users interact with their own smartphones. In addition to detecting depression, we explore the minimum number of days user data required to produce reliable performance. The following sections describe our approach in detail, starting with the design of our passively running mobile affective system.

%% file: Section/3_Method.tex
\section{Design Open-Sourced Affective Mobile System for Researchers in HCI Community}
Our framework design is built upon the HCI theory and affective computing research. This section introduces an overview of designing our affective mobile system, FacePsy (Section \ref{sec:TAF}, \ref{sec:UBS}, \ref{sec:DR}, \ref{sec:CP})., and evaluates the feasibility of the system in a pilot study to refine the FacePsy (Section \ref{sec:FS}). FacePsy is open-sourced with several key objectives. Our primary goal is to encourage widespread adoption of the system, enabling users to derive value from the generated data and facilitating engagement with topics related to mental health sensing, particularly in the context of depression. Additionally, we aim to cultivate a community of contributors who can enhance the system by designing FacePsy with modularity as a central principle. The complete source code for FacePsy can be accessed on GitHub: https://github.com/stevenshci/FacePsy.

\subsection{Technical Aspects of FacePsy}
\label{sec:TAF}
FacePsy is designed to capture real-time facial behavior primitives as users interact with their mobile devices. Operating with a response time of 2.5 Hz, which was robustly tested across two different devices, the app leverages the front camera to gather facial data during specified triggers opportunistically. This approach enhances data relevance and optimizes energy consumption and privacy. FacePsy integrates advanced modules such as facial landmark detection \cite{mlkit}, head pose estimation \cite{mlkit}, and facial action unit recognition \cite{ertugrul2019cross}, running these sophisticated processes directly on the device. This on-device processing ensures privacy and increases energy efficiency by eliminating the need for continuous data transmission. In response to the challenges prevalent in the HCI and Ubicomp domains concerning the deployment of everyday facial behavior sensing systems, FacePsy emphasizes: (1) Achieving high performance in facial image capture and feature extraction without compromising the user experience. The tested response time ensures that the app functions effectively in real-time. (2) Prioritizing on-device image processing to safeguard user privacy and improve battery efficiency. (3) Implementing trigger-based data collection to refine model performance, reducing the need for continuous data monitoring and processing. (4) Enhancing system controllability and configurability, which allows researchers to customize data collection parameters according to specific research needs.

The rest of the section describes the facial behavior primitives detection implemented in the FacePsy system to achieve the required research goal in detail. 

\subsection{Unobtrusive Background Sensing on A User's Phone}
\label{sec:UBS}
FacePsy has introduced functionalities, including configurable app triggers for data sampling. Researchers can now set data collection trigger conditions and sampling rates, with a default set to 10 seconds, informed by studies indicating peak emotional responses within this timeframe \cite{ekman2003emotions, matsumoto2011evidence}. The app supports real-time feature extraction using the smartphone's camera. We quantized a CNN model \cite{ertugrul2019cross} to integrate into our system; it efficiently extracts facial behavior features, with processing time varying based on the device's capabilities and feature complexity. Processed images are automatically discarded from the user device after 20 seconds to ensure user privacy and manage storage. None of the processed images leave the user's device or are processed outside of user's device. Additionally, our system offers researchers flexibility in defining sampling triggers and rates, enhancing its applicability for diverse research needs.

Upon installation, the mobile app registers as a background service, monitoring user events like phone lock/unlock and application usage (e.g., WhatsApp, Twitter). These events trigger a 10-second data collection session using a photo burst, a duration optimized for battery and computational efficiency based on our feasibility study. FacePsy captures facial markers, such as Action Units \cite{ekman1978facial}, and securely syncs this data to a research server during this session. To balance image processing demands and resource consumption, the app records at a rate of 2.5Hz, ensuring seamless phone usage. This decision was informed by a feasibility study involving two users. An alternative for a higher frame rate could involve recording as a media stream and processing it frame-by-frame using a codec.

\subsection{Design Rationale Behind Facial Behavior Primitives Selection}
\label{sec:DR}
Our framework design is based on principles of HCI theory and is informed by extensive research in affective computing. Theoretical Backgrounds on hand-crafted features (e.g., HOG, LBP, etc.) \cite{song2019dynamic, dhall2015temporally} or deep-learned \cite{zhou2018visually, jan2017artificial} features have adopted to represent each frame or short video segment in lab settings. However, traditional hand-crafted features are not optimal for facial behavior applications as they are not specifically designed for this purpose. Based on previous studies, which suggest that non-verbal visual cues characterize depression, our proposed approach uses facial behavior attributes such as Action Units (AUs), face landmarks, and head pose as frame-wise descriptors. The machine learning kit (ML Kit) \cite{mlkit} is used to automatically detect face landmarks, smile and eye-open probability, and head pose, facilitating data collection. A convolutional neural network (CNN) \cite{ertugrul2019cross} is adapted to detect the intensities of 12 different AUs, resulting in 151-channel facial behavior time-series data (12 AU, 1 smile probability, 2 eye open probability, 3 head pose, and 133 face landmarks) for each session. 
 
AUs from the facial action coding system (FACS) \cite{ekman1978facial}, which taxonomizes human facial movements, specifically AU4, AU12, AU15, and AU17, have been linked to depression severity \cite{song2020spectral, gavrilescu2019predicting} and mood disorders \cite{hong2019exploring, kollias2021affect}. Additionally, 133 facial landmarks that localize key facial regions are detected using ML Kit \cite{prabhu2017real}. Features extracted from these landmarks, such as Eye-aspect ratio (EAR) \cite{feng2020using} and intervector angles (IVA) \cite{islam2016sention}, have associations with hypervigilance \cite{benoit2005hypovigilence}, drowsiness \cite{maior2020real}, and facial expression analysis \cite{islam2016sention}. Moreover, they've been instrumental in assessing mental fatigue \cite{cheng2019assessment}. We also computed probabilities for smile and eye-open states, which have been linked to depression \cite{gehricke2000reduced} and fatigue \cite{zhang2017driver, kroencke2000fatigue}. Lastly, our app captured head Euler angles representing head movements. Head pose features, such as slower head movements and specific head orientations, have been identified as indicators of depression \cite{song2020spectral, alghowinem_head_2013} and suicidal ideation \cite{laksana2017investigating, eigbe2018toward}. 
 
 The rationale behind selecting these features is rooted in their established associations with mental health indicators, especially depression. By integrating a wide range of facial attributes, our approach aims to provide a holistic and unbiased representation of facial behaviors. Compared to previously employed hand-crafted and deep-learned features, the facial behavior descriptors provided in this work have several advantages. They are impartial since their values are unrelated to the subjects' identities, which prevents bias based on gender, age, ethnicity, etc., from influencing the results, as suggested by Song et al. \cite{song_spectral_2020}. Second, They have a clear, comprehensible meaning, which makes them more interpretable.

\subsection{Configurability, Privacy and User Awareness}
\label{sec:CP}
To provide high configurability for tailoring the application as needed, the FacePsy helps researchers adjust the duration of data collection, which by default is set to 10 seconds upon any trigger for data collection. Furthermore, it allows for distinct data collection durations based on different trigger types, such as app usage, and phone unlocking. Lastly, the system supports the configuration of various app usage triggers to initiate data collection, ensuring a comprehensive and adaptable research tool.

 \begin{figure}[h]
    \includegraphics[scale=0.27]{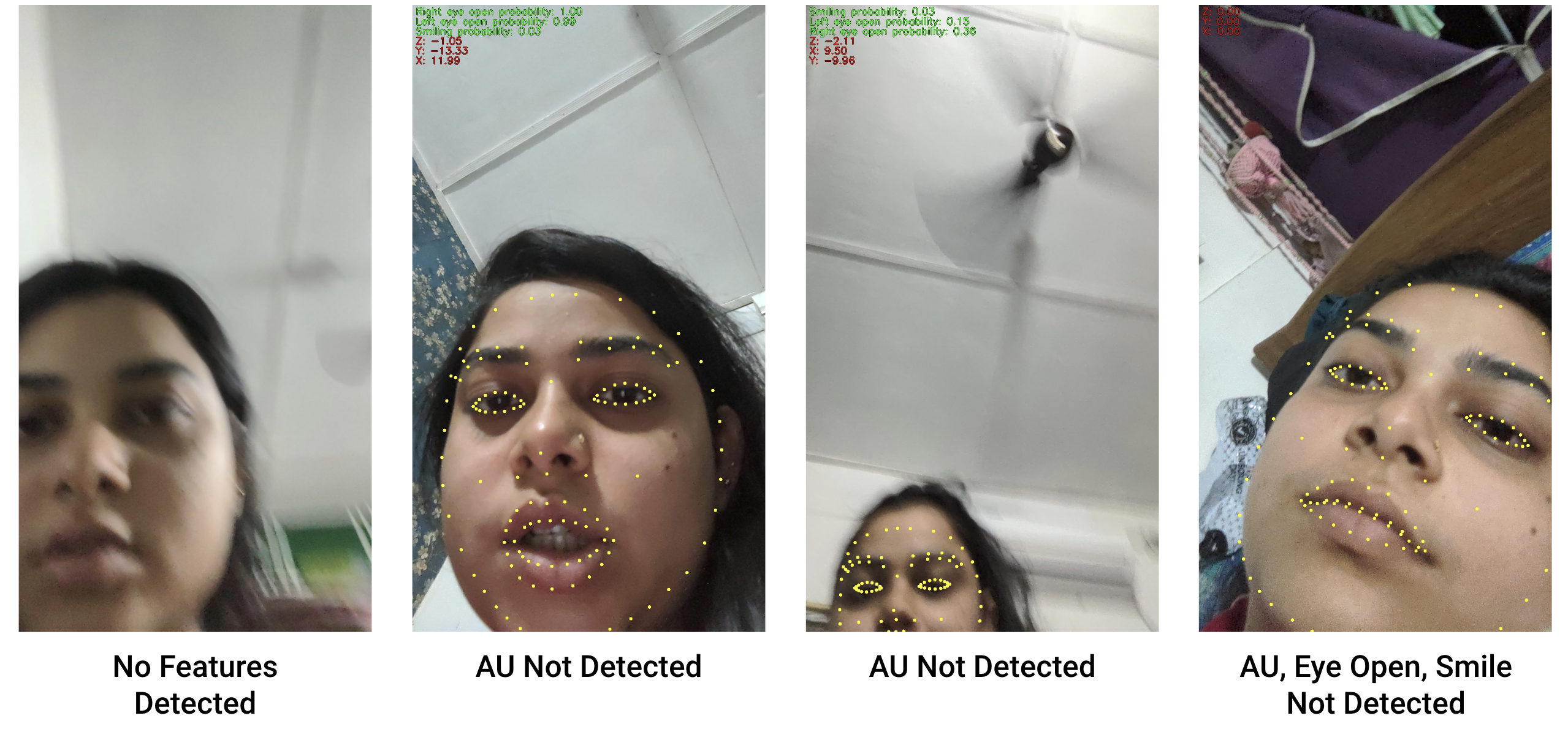}
    \caption{Examples of Test Images that Fails to Capture Features}
    \label{fig:FacePsyFail}
\end{figure}

When designing our system, paramount importance was given to user privacy and awareness. Drawing inspiration from HCI research, we investigate user nudging \cite{balebako2014improving, felt2012ve} in privacy systems \cite{denning2014situ}. This involves giving information about background data collection, especially in private contexts where no images are stored. Firstly, a clear notification is displayed in the notification bar, indicating "Data collection is active" to ensure users are always aware when data is being collected. Additionally, a green light indicator is positioned next to the camera, signaling when the camera is actively capturing data. In the event FacePsy app automatically restarts itself in the background, either due to a phone restart or an app crash, a toast notification is presented to the user, stating "FacePsy is running on background" Lastly, to further safeguard user data, once facial behavior primitives are extracted, the processed images are automatically deleted from the user's device.

\begin{figure}[h]
    \includegraphics[scale=0.27]{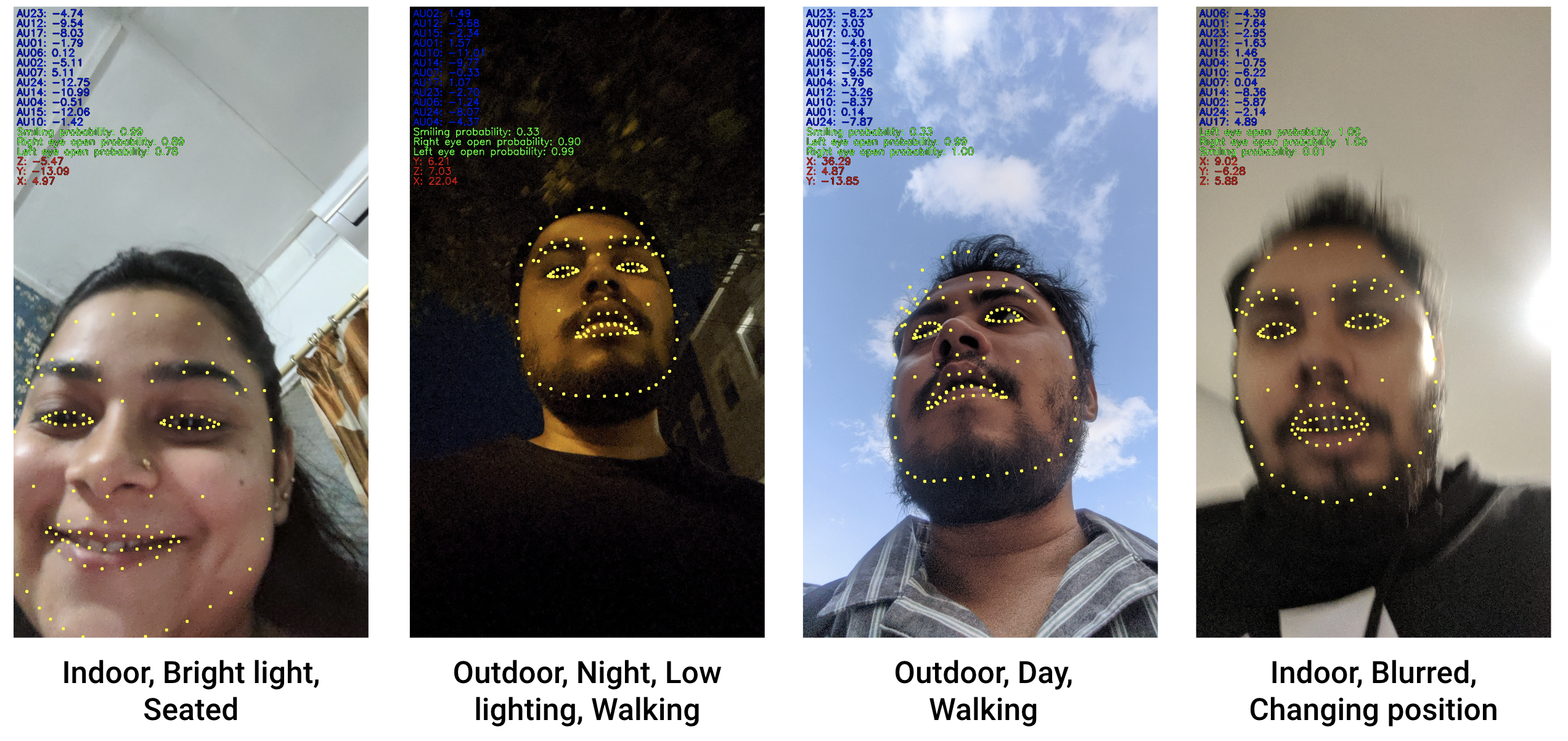}
    \caption{Examples of Test Images That Succeeds to Capture Features }
    \label{fig:FacePsySuccess}
\end{figure}

\begin{table}[h]
\centering
\small
\caption{FacePsy App Resource Usage on Google 4 \& 5a}
\begin{tabular}{lccc}
\toprule
Resource & Day 1 (T1, T2) & Day 2 & Avg (T1 \& T2) \\
\midrule
Battery & 37\%, 57\% & 58\%, 43\% & 48.75\% \\
Memory (MB) & 133, 144 & 85, 57  & 104.75 \\
Data usage (MB) & 9.25, 22.23 & 16.13, 6.3 & 13.48 \\
Storage (MB) & 175, 155 & 177, 177 & 171 \\
\bottomrule
\label{tabs:RU}
\end{tabular}
\end{table}

\subsection{Feasibility Study}
\label{sec:FS}
Prior to investigating if facial behavior primitives collected from FacePsy can assess a user's depression status, we decided to carry out a feasibility study to see whether the system could precisely detect the user's facial behaviors using the front-facing camera of a smartphone. To do this, we enlisted the help of 1 volunteer (T2: tester 2) and the first author (T1: tester 1), who used a Google Pixel 4 and 5a smartphone to gather data for two days. Participants carried FacePsy installed in their phones into their daily lives. FacePsy collected data whenever participants unlocked their phones. We evaluate our system for perceived slowness in their device, causing any delay, interruption, or disruption to phone usage. Our facial behavior sensing modules generated pictures annotated action units, smile probability, eye open probability, head pose, and face landmarks. After gathering these pictures, the first author confirmed if the facial behavior primitives were correctly detected. To do this, the first author manually verified that the annotated photographs matched the unannotated photos. Annotated images of facial behavior markers that were identified are shown in Figures \ref{fig:FacePsySuccess}, while images were accurate, the processing modules made errors on several occasions (See Figure \ref{fig:FacePsyFail}). On average, the app consumed 48.75\% battery while running in the background and 104.75mb memory. See Table \ref{tabs:RU} for more details. In total, we collected 834 images. Our features extraction module processed only 817 images where a face was detected with an average failure rate of 2.04\% (See Table \ref{tab:FacePsyPerfromance}). This was calculated based on daily end-of-day reports from volunteers regarding resource consumption by our app. To optimize resource consumption, we implement opportunistic data collection, which allows the app to collect data less frequently, thus reducing the load on device resources. Further refining data collection triggers can be done to ensure that the app only collects data when optimal behavioral signals are present, reducing unnecessary resource usage.

\begin{table}[h]
\centering
\small
\caption{FacePsy Data Processing Performance}
\begin{tabular}{lcc}
\toprule
\textbf{Metric} & \textbf{T1} & \textbf{T2} \\
\midrule
Total Images & 411 & 423 \\

Successful Extractions & 401 & 416 \\

Number of Failures & 10 & 7 \\

Overall Failure Rate (\%) & 2.43\% & 1.65\% \\

Failure Rate for AU (\%) & 10.97\% & 24.28\% \\

Failure Rate for Landmarks (\%) & 0.00\% & 0.00\% \\

Failure Rate for Classification(Eye, Smile) & 0.00\% & 1.68\% \\

Failure Rate for Head Pose (\%) & 0.00\% & 0.00\% \\
\bottomrule
\end{tabular}

\label{tab:FacePsyPerfromance}
\end{table}

The results of our app feasibility evaluation were encouraging. Our system detected the facial behavior makers without hindering the user device experience at a 10-second photo burst in the background.
\nopagebreak

\section{Field Study Data Collection}
In this section, we describe our study protocol and data collection process in naturalistic environments.

\subsection{Participants}

Of N=25 participants (mean age 27.88 $\pm$ 8.87, range 18 - 48)\footnote{9 participants did not provide an age in the demographic survey.}, 8 were females, 11 were males, and 6 did not specify gender on the demographic survey. 15 participants were Asian, 4 participants were Caucasian, and 6 participants did not specify their ethnicity on the demographic survey. 1 participant indicated that their highest education was high school, 8 participants had bachelor's degrees, 10 participants had master's degrees, and 6 participants did not indicate their highest education on the demographic survey. 4 participants indicated that they had been diagnosed with a mental disorder in the past, 15 indicated they had not, and 6 did not answer the question on the demographic survey. Detailed demographic distribution is provided in Table \ref{tab:demographic}.

\begin{table}
\centering
\small
\caption{\label{tab:demographic} Demographic Distribution}  
\setlength{\tabcolsep}{4pt}  
\begin{tabular}{lcccc}
\toprule
\textbf{Attribute} & \textbf{Unspecified} & \textbf{Male} & \textbf{Female} & \textbf{Total} \\
\midrule
\textbf{Gender} & 6 & 11 & 8 & 25 \\
\textbf{Age (Average)} & - & 24.11 & 32.71 & 27.88 \\
\textbf{Ethnicity} &  &  &  &  \\
- Unspecified & 6 & 0 & 0 & 6 \\
- Asian & 0 & 9 & 6 & 15 \\
- Caucasian & 0 & 2 & 2 & 4 \\
\textbf{Education} &  &  &  &  \\
- Unspecified & 6 & 0 & 0 & 6 \\
- High School & 0 & 0 & 1 & 1 \\
- Bachelor's Degree & 0 & 5 & 3 & 8 \\
- Master's Degree & 0 & 6 & 4 & 10 \\
\textbf{Mental Health Rate (Average on a scale of 1-10)} & - & 6.73 & 6.88 & 6.79 \\
\textbf{Depression State} &   &   &   &   \\
- Unspecified & 6 & 0 & 0 & 6 \\
- Not at all often & 0 & 2 & 1 & 3 \\
- Not so often & 0 & 4 & 4 & 8 \\
- Somewhat often & 0 & 5 & 1 & 6 \\
- Very often & 0 & 0 & 2 & 2 \\
\textbf{Mental Disorder Diagnosis} &  &  &  &  \\
- Unspecified & 6 & 0 & 0 & 6 \\
- No & 0 & 9 & 6 & 15 \\
- Yes & 0 & 2 & 2 & 4 \\
\textbf{Smoking Marijuana} &  &  &  &  \\
- Unspecified & 6 & 0 & 0 & 6 \\ 
- No & 0 & 9 & 8 & 17 \\ 
- Yes & 0 & 2 & 0 & 2 \\
\bottomrule
\end{tabular}
\end{table}

\subsection{Participants and Study Procedure}
This study was reviewed and approved by the Institutional Review Board (IRB) at the University. Participants in this study were on-boarded remotely across multiple time zones via Zoom Conference Meetings. Participants were eligible to participate in the study if they were above 18 and owned a data plan-enabled Android smartphone. The research team advertised the study through flyers and posts on Facebook and Whatsapp groups. Participants were asked to respond to a screening questionnaire and select a preferred time for the onboarding Zoom meeting. In the onboarding meeting, the interviewer gave participants informed consent and asked them to respond to the baseline questionnaire. After the baseline, the interviewer took a semi-structured interview to understand the participant's mental health, followed by installing a mobile application on the participant's device to track sensor data from their smartphones. The study questionnaires were delivered through email and administered with Qualtrics, an online survey platform.

Out of 38 participants who were initially recruited, only 25 participants completed the study. One participant reported having a high battery drain because of our app and dropped out of the study after two days. We later found the participant had high social media usage, which resulted in frequent data collection triggers. Among others, 3 participants dropped out for personal reasons, 5 didn't complete surveys, and 4 had incompatible Android versions, leading to the failure of the data collection trigger module. The participants were compensated up to \$135 for full compliance with the study. The participants were compensated \$20 for baseline and installing the data collection app and were compensated \$25 weekly for 4 weeks.

\subsection{Ground-Truth: Mental health measures} \label{Sec:GT_MHM}
Participants' depression symptoms were assessed using a self-reported 9-item Patient Health Questionnaire-9 (PHQ-9) \cite{kroenke2001phq} at three distinct times: upon joining the study (baseline), two weeks into the study (mid-point), and at the conclusion of the study (end-point). Each item on the PHQ-9 is scored from 0 (not at all) to 3 (nearly every day), with the total scores ranging from 0 to 27. This scoring captures the frequency of symptoms such as mood, sleep issues, fatigue, and changes in appetite over the past two weeks. Monitoring a single person's multiple PHQ-9 scores over time can yield valuable insights into their mental health progression. The PHQ-9 categorizes depression severity into five levels: scores of 0–4 signify no depression symptoms, 5–9 indicate mild depressive symptoms, 10–14 represent moderate depressive symptoms, 15–19 signify moderately severe depressive symptoms, and 20–27 denote severe depressive symptoms. This method allows researchers and clinicians to track the severity and changes in depressive symptoms effectively.

We can define a depressive episode as a period characterized by persistent feelings of sadness, hopelessness, and a lack of interest or pleasure in most activities observed with a PHQ-9 score during a two-week observation period. we label two weeks of data from the participant as depressed or non-depressed (i.e., a depressive episode) based on the PHQ-9 score of the participant at the beginning and end of the two-week observation period, which falls within the range of mild depressive symptoms or worse according to the PHQ-9 severity scale \cite{gilbody2007screening}. We label an individual as having a depressive episode only if the PHQ-9 score of the person is equal to or greater than 5 at both the beginning and end of the observation period; otherwise, it is considered a non-depressive period. This approach ensures consistency in labeling periods as depressive or non-depressive based on established thresholds of depressive symptom severity. We complement our binary classification models by incorporating regression models designed to predict the PHQ-9 scores. It’s important to note that the PHQ is a versatile instrument, used both for screening for depression and for monitoring changes in clinical symptoms \cite{kroenke2010patient}. In total, we label 14 cases of depressive episodes and 30 non-depressive episodes (where depressive episode length is two weeks). We excluded the last two weeks of data from 6 participants due to non-compliance, resulting in the exclusion of 6 depressive episodes.

\subsection{Facial behavior data collection}
The FacePsy app activates to capture facial data in three circumstances: when the user unlocks their phone when the user accesses one of a preset number of trigger apps. The phone unlock trigger activates the FacePsy app for 10 seconds after a user unlocks their phone. FacePsy also activates for 10 seconds when the user opens one of thirty-five different apps in total, divided into the categories of communication, social, productivity, entertainment, and health. For example, Instagram, Google Chrome, and Android Messages are all considered trigger apps. The images processed were mostly clear and had high resolution. However, some images had some noise or blurriness on which the model could not detect faces, which is a precursor for routines such as AU detection, landmark detection, etc. These frames were dropped.

\section{Data Processing And Analysis}
\subsection{Feature Engineering} \label{Section-Feature-Extraction}
Since most of the features are extracted by our behavior-sensing system on the user phone itself, we extract very few features as part of post-processing. Our system pre-extracts features such as Action Units (AU1, 2, 4, 6-7, 10, 12, 14, 17, 23-24), Smiling and Eye open probability, face landmarks, and Head pose features (yaw, pitch and roll) on user device itself. We additionally extract features such as the Eye-aspect ratio and Inter-vector angles. More details on these features are described below. 

\subsubsection*{Inter-vector angle}
Inter-Vector Angles (or IVA) are scale-invariant geometric features computed on facial landmarks for the purpose of facial shape representation \cite{islam2016sention}. We consider the nose center as the centroid of the face for the purposes of computing IVA features. We then segment the face into 8 regions (nose center, jawline, left eyebrow, left eye, right eyebrow, right eyebrow, mouth, and cheeks) and compute in total 1439 triangles by taking permutations of all possible triangles from the centroid to the remaining facial landmarks. We then use Principle Component Analysis to reduce the number of IVA features down to 10. We then compute angular velocity and acceleration. 

\subsubsection*{Eye-aspect ratio}
Eye Aspect Ratio is a measure of the aspect ratio of the eye region, which is used as an estimate of the eye-opening state. We defined EAR as the sum of two vertical lengths of the eye divided by two times the horizontal length of the eye. 

We collected 12 Action Units (AU), 1 Smile Probability, 2 Eye Open Probabilities, 3 Head Euler Angles, 2 Eye Aspect Ratios, and 20 Inter-Vector Angles. We segmented each participant's day into four epochs: midnight (12am-6am), morning (6am-12pm), afternoon (12pm-6pm), and evening (6pm-12am), each lasting six hours. For each epoch, we computed statistical features such as min, max, mean, median, sum, std, q1 and q3 to summarize the features of that epoch. After this procedure, we have 320 features for each epoch in our final dataset. We then classified each instance into its respective depression class. We noticed variations in participation duration, with an average of 3.36 days missing due to participants' early exit from the study. This resulted in a total dataset of 616 participant days. In total, we gathered 544 days of facial data from 25 participants over a four-week period. Notably, there were 55 days without any recorded data, as explained by participants, such as planned holidays or breaks. A further 17 days lacked data due to issues like image quality and eye-open probability in the feature extraction process. As a result, the total number of effective data points for analysis is 544.

\subsection{Statistical Analysis}
The primary target variable of interest in our statistical analysis was the presence or absence of a depressive episode. We calculated Pearson's correlation coefficient (r-value) for each feature within our dataset to assess its relationship with the target variable. Additionally, we determined the mean and standard deviation for each feature, separating the data by group. Features were then ordered according to the absolute value of their r-values to pinpoint those with at least a weak correlation (r-value >= |0.20|). This approach allowed us to focus on the most significant relationships, improving the interpretability and efficiency of our models, minimizing unnecessary complexity, and reducing the likelihood of overfitting.

\begin{figure}
    \includegraphics[scale=0.233]{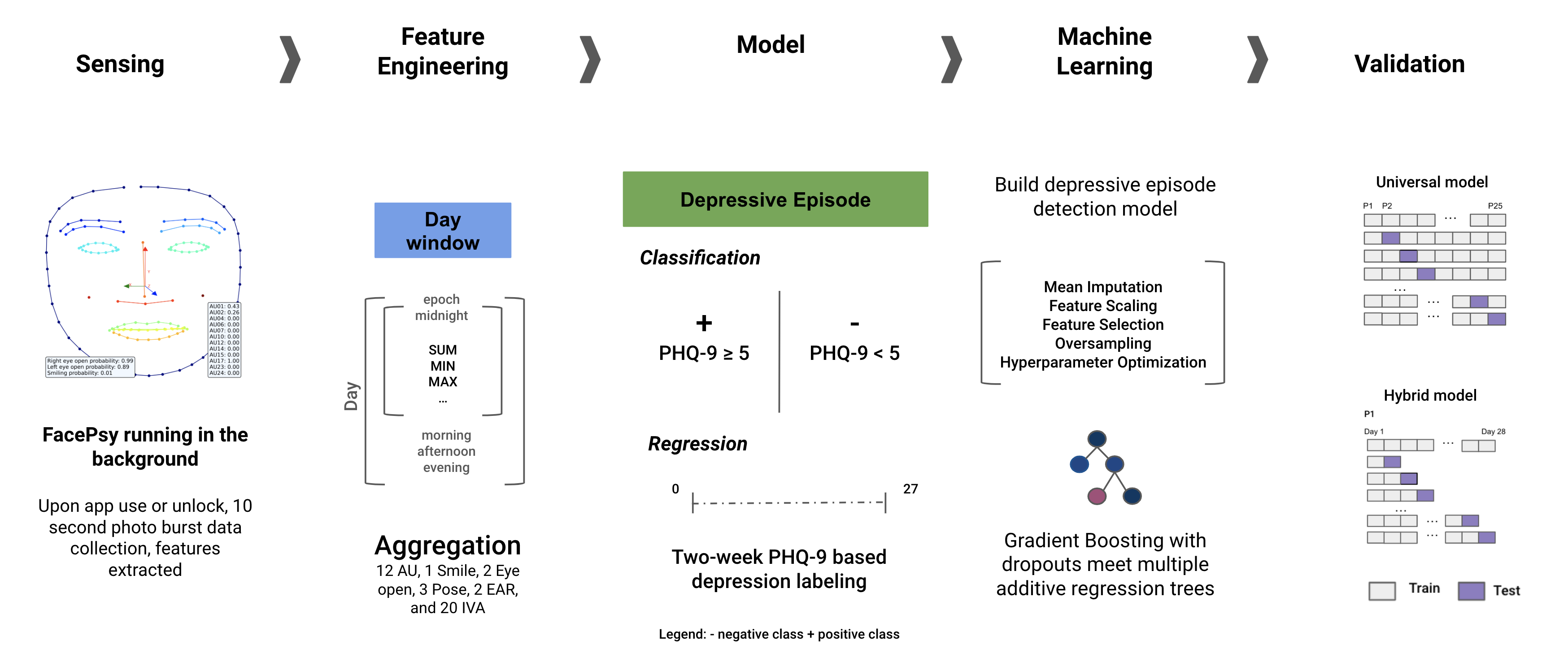}
    \caption{Overview of Our Affective Mobile System}
    \label{fig:FacePsyStudyFlow}
\end{figure}

\subsection{Feature Selection}
Our analysis used feature selection (FS) with a Decision Tree classifier, specifically the CART (Classification and Regression Trees) algorithm implemented in scikit-learn's DecisionTreeClassifier. To compute the importance scores for all features in the dataset, we used the Gini importance, a measure derived directly from the Decision Tree itself. We then established a threshold for feature selection based on the mean value of these importance scores, which calculated to 0.00078125. Features with importance values above this threshold were considered significant and retained for further analysis, while those below were discarded. This approach ensures a data-driven, objective criterion for feature selection, enhancing model interpretiveness and efficiency by focusing on features that contribute to the prediction of depressive episodes.

\subsection{Predictive Modeling with Machine Learning}
In our predictive analysis using machine learning (See Figure \ref{fig:FacePsyStudyFlow}), we developed a classification model to detect instances of depression and non-depression, as well as a regression model to predict the PHQ-9 score (See Section \ref{Sec:GT_MHM}). We assigned labels of depression and non-depression to each day of data based on their corresponding PHQ-9 scores, which served as the ground-truth values.

Our predictive modeling framework utilizes LightGBM (LGBM), a machine-learning library that implements a gradient-boosting algorithm. This method has demonstrated robust predictive capabilities in previous research \cite{opoku2022mood, opoku2021predicting} focused on depression prediction. Our goal is to develop both a universal model that identifies general patterns and a hybrid model that captures intricate interactions and temporal sequences within the data. To address the issue of class imbalance, where depressive instances are less frequent, we applied SMOTE \cite{chawla2002smote} on the training dataset to enhance the representation of the minority class, used mean imputation for handling missing data, and performed standard scaling on the features. Hyperparameter tuning was conducted tomaximize the AUROC value for classification and MAE for regression. We constructed nine supervised models, each tailored to different learning schemes and utilizing subsets of facial behavior features sourced from various facial regions to assess their predictive power for depression. The evaluation of these models primarily relies on the AUROC score \cite{huang2005using}. This metric is particularly effective for depression prediction as it evaluates a model's ability to distinguish between depressive and non-depressive states by considering both the true positive rate (TPR) and false positive rate (FPR). The insensitivity of AUROC to class imbalance makes it especially valuable, and a higher AUROC score signifies superior model performance.

\subsubsection{Universal model}
This learning scheme utilizes a standardized procedure where a single model is created for all users to identify depressive episodes. It uses a leave-one-participant-out (LOPO), a.k.a leave-one-out/leave-one-group-out cross-validation technique. This approach, commonly used in numerous mobile inference systems, provides a clear understanding of model generalizability. Once this universal model is in place, it remains unchanged.

\subsubsection{Hybrid model}
The ideal model would blend the high precision of individualized models with the ease of use of universal models that don't require user training. Our study was unable to use individualized models due to a lack of data - only two labels per participant, which is not enough for such intricate modeling. We experimented with a hybrid model, incorporating a small quantity of user-specific data with a broader general user dataset. we implement nested cross-validation to minimize the likelihood of model overfitting by implementing a robust ML model training strategy recommended by Asare et al. \cite{opoku2022mood} for the predictive analysis of depression.  We employed stratified three-fold cross-validation with a time-series aware leave-one-participant-day-out (LOPDO) cross-validation for the outer and inner cross-validation. In other words, one participant's day is chosen as the test set, and the remaining participant's dataset is chosen as the training set for each iteration of the nested cross-validation. All training set samples captured after the test set are subtracted for time-series awareness. This approach could avoid the unworkable situation in which future datasets are used to forecast the past. Consequently, the LOPDO model effectively combines unique aspects of how an individual's facial behavior data is linked to their state of depression while also identifying general patterns consistently seen across different people. The classifiers' hyperparameter optimization, feature scaling, oversampling, and missing data imputation were all addressed by inner cross-validation. Using grid search across a predetermined set of parameters, we optimized the classifiers' hyperparameters by maximizing the model AUROC score.

%% file: Section/4_Results.tex
\section{Field Study: Understanding and Detecting Depressive Episodes}
To obtain the feasibility of our proposed FacePsy framework in the field study, we address the following question: (RQ1) can the facial behavior features collected by our mobile sensing system be effectively utilized to detect depressive episodes in a naturalistic environment? (Section \ref{RQ1}). To understand sets of depression-related biomarkers that could be used in the wild we address (RQ2) What’s the significance of different biomarkers on depression detection differentiating depressive vs. non-depressive episodes in real-world settings? We introduce the top 27 facial behavior features.

\begin{table}[h]
\centering
\small
\caption{\label{tab:corr}Summary of Feature Correlations with Depressive Episodes}  
\setlength{\tabcolsep}{4pt}  
\begin{tabular}{lp{1cm}rp{2.5cm}p{3.1cm}}
\toprule
Feature & p-value (<0.05) & r-value & Depressive Episode Mean (SD) & Non-Depressive Episode Mean (SD) \\
\midrule
                 ear\_right\_sum\_morning &                   0.00 &     0.35 &                23.34 (28.87) &                      8.26 (11.1) \\
                  ear\_left\_sum\_morning &                   0.00 &     0.34 &                21.36 (25.82) &                     7.99 (10.83) \\
          headEulerAngle\_Y\_sum\_morning &                   0.00 &    -0.33 &             -404.03 (719.02) &                  -33.33 (332.65) \\
    leftEyeOpenProbability\_sum\_morning &                   0.00 &     0.33 &                55.41 (69.64) &                     21.04 (28.1) \\
   rightEyeOpenProbability\_sum\_morning &                   0.00 &     0.31 &                 48.74 (62.6) &                    19.99 (25.73) \\
                    AU15\_min\_afternoon &                   0.00 &     0.27 &                  0.09 (0.14) &                      0.04 (0.04) \\
   rightEyeOpenProbability\_std\_evening &                   0.00 &     0.26 &                  0.24 (0.07) &                       0.2 (0.07) \\
        smilingProbability\_sum\_morning &                   0.00 &     0.26 &                  5.52 (9.62) &                      1.98 (3.52) \\
 rightEyeOpenProbability\_std\_afternoon &                   0.00 &     0.23 &                  0.24 (0.07) &                       0.2 (0.07) \\
                     AU17\_sum\_midnight &                   0.00 &    -0.23 &                  6.36 (13.6) &                    21.02 (34.59) \\
                   AU12\_median\_morning &                   0.00 &    -0.22 &                  0.29 (0.21) &                       0.4 (0.27) \\
                      AU07\_std\_evening &                   0.00 &     0.22 &                  0.26 (0.08) &                      0.22 (0.07) \\
                      AU02\_std\_evening &                   0.00 &     0.21 &                  0.21 (0.07) &                      0.17 (0.08) \\
        smilingProbability\_max\_evening &                   0.00 &     0.21 &                   0.35 (0.2) &                      0.25 (0.21) \\
                   AU07\_median\_morning &                   0.00 &    -0.21 &                  0.59 (0.24) &                       0.68 (0.2) \\
        smilingProbability\_max\_morning &                   0.00 &     0.21 &                   0.33 (0.2) &                       0.24 (0.2) \\
      smilingProbability\_mean\_midnight &                   0.00 &     0.21 &                  0.14 (0.12) &                      0.09 (0.09) \\
                       AU12\_q3\_morning &                   0.00 &    -0.21 &                  0.39 (0.23) &                       0.5 (0.27) \\
                     AU12\_mean\_morning &                   0.00 &    -0.21 &                  0.31 (0.19) &                      0.41 (0.24) \\
    leftEyeOpenProbability\_std\_evening &                   0.00 &     0.20 &                  0.23 (0.07) &                       0.2 (0.07) \\
          headEulerAngle\_X\_std\_evening &                   0.00 &     0.20 &                   3.86 (1.4) &                      3.25 (1.33) \\
                      AU06\_max\_morning &                   0.01 &    -0.20 &                  0.56 (0.27) &                      0.66 (0.23) \\
        smilingProbability\_std\_evening &                   0.00 &     0.20 &                  0.09 (0.05) &                      0.06 (0.06) \\
      smilingProbability\_std\_afternoon &                   0.00 &     0.20 &                  0.08 (0.06) &                      0.06 (0.06) \\
     smilingProbability\_mean\_afternoon &                   0.00 &     0.20 &                  0.11 (0.09) &                      0.07 (0.09) \\
                     AU06\_mean\_morning &                   0.01 &    -0.20 &                  0.29 (0.19) &                      0.38 (0.21) \\
       smilingProbability\_mean\_evening &                   0.00 &     0.20 &                   0.1 (0.08) &                      0.07 (0.07) \\
\bottomrule
\end{tabular}
\end{table}

\subsection{Statistical difference between depressive and non-depressive episodes}

The analysis evaluated the correlation of various features with depressive episodes (See Table \ref{tab:corr}). The features were ranked based on the strength of their correlation (r-value) with the target variable, which indicates the presence of a depressive episode. Out of 1280 only 158 features had a p-value less than 0.05. Selecting features with at least a weak correlation (r-value >= abs(0.20)), streamlining analysis by prioritizing meaningful relationships, and enhancing model interpretability and efficiency while reducing noise and the risk of overfitting. Temporal dynamics of depressive episodes suggest features measured in the morning often show significant correlations, pointing to the potential impact of depression on morning routines or states, such as reduced facial expressiveness or specific eye movement patterns. The feature \textit{headEulerAngle\_Y\_sum\_morning} has the strongest negative correlation with depressive episodes, with an r-value of -0.33. This suggests that as the value of this feature decreases, the likelihood of a depressive episode increases. The Y rotation of the head translates to the yaw of the Euler angle. Other features with notable negative correlations include \textit{AU17\_sum\_midnight}, \textit{AU12\_median\_morning}, \textit{AU12\_q3\_morning} \textit{AU07\_median\_morning}, \textit{AU06\_max\_morning} and \textit{AU06\_mean\_morning}. The absence of \textit{AU06}, associated with expressing emotions related to happiness or joy, correlates with depression. Furthermore, \textit{AU07}, \textit{AU12} and \textit{AU17} have been linked to depression severity, supporting existing evidence \cite{song2020spectral, gavrilescu2019predicting}.

The feature \textit{ear\_right\_sum\_morning} and \textit{ear\_left\_sum\_morning} shows a strong positive correlation with depressive episodes, with an r-value of 0.35 and 0.34, respectively. This indicates that as the value of this feature increases, the likelihood of a depressive episode also increases. It's very important to consider the temporal dynamics of these features. Other features with significant positive correlations include \textit{leftEyeOpenProbability\_sum\_morning}, \textit{rightEyeOpenProbability\_sum\_morning}, \textit{rightEyeOpenProbability\_std\_evening} and \textit{leftEyeOpenProbability\_std\_evening}. The presence of strong EAR, eye open probability related to high alertness \cite{abe2023perclos} in morning and evening could be explained as the “eveningness–morningness” dimension in depression \cite{chelminski1999analysis}. The preference for morning or evening can largely be attributed to the reduction of depressive symptoms such as low energy, avoidance of social interaction, and loss of interest in previously pleasurable activities \cite{putilov2017state}. 

The presence of a positive correlation (r-values of 0.26, 0.21 for morning sum and maximum, respectively; and similarly positive correlations for evening and midnight measures) between \textit{smilingProbability} and depressive episodes suggests that higher smiling probabilities are associated with an increased likelihood of depressive episodes. This interpretation may seem counterintuitive since it's expected that depressive episodes would be associated with less smiling. However, this unexpected positive correlation doesn't necessarily imply that smiling more leads to depression or vice versa. It might reflect complex underlying behaviors or compensatory mechanisms, such as "smiling depression," where individuals might smile or maintain a facade of happiness in social situations despite experiencing depressive symptoms internally \cite{vanswearingen1999specific}. However, as a limitation of our study we are not able to confirm if participants are going such cases. Previous research suggests that depression is not only associated with sad facial expressions but also with “a total lack of facial expression corresponding to the lack of affective experience” \cite{ellgring2007non}. Since we collect short segments (10 sec) of data, it can be interpreted in various ways, e.g., a smile may be a result of feeling happy or feeling helpless, as suggested by prior research \cite{song2020spectral}. While a positive correlation between \textit{smilingProbability} and depressive episodes seems paradoxical, it highlights the complexity of depressive behaviors and the importance of considering broader psychological and situational contexts when interpreting these findings.

\subsection{Model development from data in the field study } \label{RQ1}

\subsubsection{Universal model}
\begin{figure}[h]
    \includegraphics[scale=0.3]{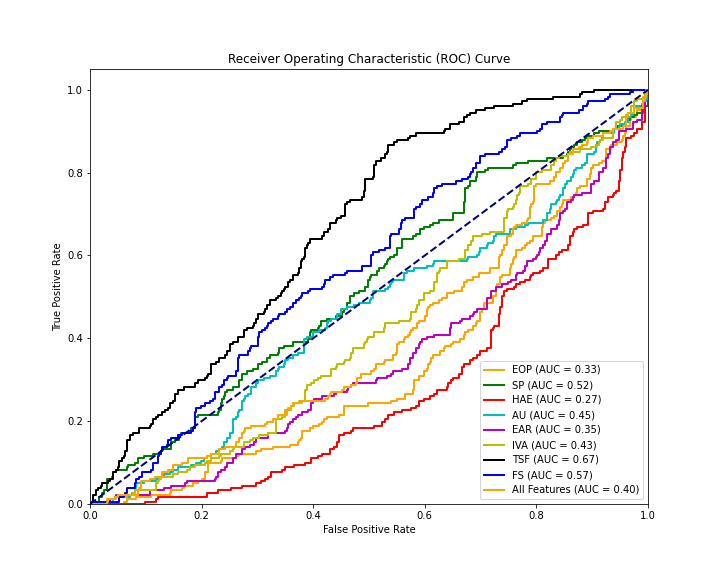}
    \caption{The ROC plots show the universal model performance of each feature type model.}
    \label{fig:roc_features_lopo_lgbm}
\end{figure}

Table \ref{tab:results_features_lopo_lgbm} summarizes the predictive performance of universal models. To understand how different subsets of facial behavior features contribute to detecting depression, we evaluated nine different models, each with a different face feature set, using LightGBM. The model using the most significant features from the correlation analysis performed the best, followed by the model that included feature selection. This approach enhances the model's interpretability and comprehension. The TSF model achieved 51\% accuracy, with a precision of 40\%, indicating that it correctly predicted depression 40\% of the time. A recall of 96\% suggests that out of all the depressive episode cases in the dataset, the model successfully identifies 96\% of them as positive. This is particularly important in depression detection, where missing out on positive cases leads to missing out on opportunities to intervene. The model's reliability is also reflected in an AUROC score of 0.67 (Fig. \ref{fig:roc_features_lopo_lgbm}). In terms of regression metrics, the MAE for each model also provides insights into the quantitative accuracy of depression severity estimation, showing the lowest error (3.26) for the Action Units model, which suggests its superior ability to estimate the severity of depression correctly compared to other models, where TSF model got an MAE of 5.13. While this indicates limited ability to distinguish between depression and no-depression classes, it represents better agreement between the model's predictions and actual observations than a random classifier.

\begin{table}[h]
\caption{\label{tab:results_features_lopo_lgbm} Universal Model Performance: We trained eight LGBM models for predicting depression, including a different feature subset. The model trained using all features showed the best results in predicting depression} 

\footnotesize
\centering  
    \begin{tabular}{p{3.4cm}p{1cm}p{1cm}p{1cm}p{1cm}p{1cm}p{1cm}p{1.7cm}}
    \toprule
    \textbf{Model} & \textbf{MAE} & \textbf{Accuracy} & \textbf{Precision} & \textbf{Recall} & \textbf{F1} & \textbf{AUROC} & \textbf{No. of Features} \\ 
    \midrule
    
    Eye Open Probability (EOP)  & 5.20 & 0.33\ & 0.31\ & 0.83\ & 0.45\ & 0.33\  &  64 \\
    
    Smiling Probability (SP) & 5.32 & 0.37\ & 0.33\ & 0.87\ & 0.48\ & 0.52\ & 32 \\
    
    Head Euler Angle (HEA) & 4.71 & 0.29\ & 0.28\ & 0.71\ & 0.40\ & 0.27\ & 96 \\

    Action Units (AU) & \textbf{3.26} & 0.38\ & 0.30\ & 0.66\ & 0.42\ & 0.45\ & 384 \\

    Eye-aspect ratio (EAR) & 5.31 & 0.32\ & 0.30\ & 0.80\ & 0.44\ & 0.35\ & 64 \\

    Inter-vector angle (IVA) & 4.59 & 0.40\ & 0.31\ & 0.66\ & 0.42\ & 0.43\ &  640 \\
    
    Top Significant Features (TSF) & 5.13 & \textbf{0.51}\ & \textbf{0.40}\ & \textbf{0.96}\ & \textbf{0.56}\ & \textbf{0.67}\ & 27 \\

    Feature Selection (FS) & 4.04 & 0.50\ & 0.38\ & 0.77\ & 0.51\ & 0.57\ &  46 \\
    
    All features & 3.77 & 0.40\ & 0.28\ & 0.51\ & 0.36\ & 0.40\ & 1280 \\
    
    \bottomrule
 
\end{tabular}
\end{table}

\subsubsection{Hybrid model}
Table \ref{tab:results_features_lopdo} summarizes the predictive performance of hybrid models. The model with the best performance is the one using feature selection with LightGBM. In the context of detecting depressive episodes, the model demonstrated a commendable performance with an accuracy of 69\%. Notably, when predicting a depressive episode, it was correct 57\% of the time, as indicated by a precision for the depressive class. Furthermore, it successfully identified 62\% of all actual depressive episodes, reflected by a recall. The F1-score, a measure of the model's balance between precision and recall, was 0.67 for depressive episodes, suggesting a harmonized performance despite the inherent class imbalance. The model's reliability was also underscored by an AUROC of 0.81, indicating a strong ability to distinguish between the two classes and a good agreement between the model's predictions and actual observations. The regression results further enhance our understanding, with the model achieving an MAE of 3.08 on the PHQ-9 scale, which ranges from 0 to 27. This indicates that the model's depression severity predictions are typically within approximately three points of the actual clinical assessments, showing relatively moderate accuracy in quantifying the severity of depressive symptoms.

\begin{figure}[h]
    \includegraphics[scale=0.3]{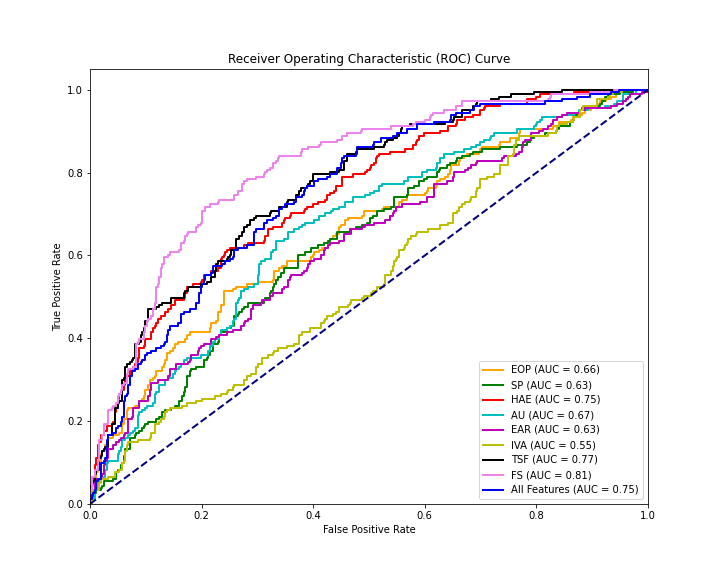}
    \caption{The ROC plots show the hybrid model performance of each feature type model.}
    \label{fig:roc_features_lopdo_lgbm}
\end{figure}

\begin{table}[h]
\caption{\label{tab:results_features_lopdo} Hybrid Model Performance : We trained eight LGBM models for predicting depression, including a different feature subset. The model trained using all features showed the best results in predicting depression} 

\footnotesize
\centering  
    \begin{tabular}{p{3.4cm}p{1cm}p{1cm}p{1cm}p{1cm}p{1cm}p{1cm}p{1.7cm}}
    \toprule
    \textbf{Model} & \textbf{MAE} & \textbf{Accuracy} & \textbf{Precision} & \textbf{Recall} & \textbf{F1} & \textbf{AUROC} & \textbf{No. of Features} \\ 
    \midrule
    
    Eye Open Probability (EOP) & 3.16 & 0.67\ & 0.50\ & 0.48\ & 0.49\ & 0.66\  &  64 \\

    Smiling Probability (SP) & 3.26 & 0.64\ & 0.46\ & 0.43\ & 0.44\ & 0.63\ &  32 \\
    
    Head Euler Angle (HEA) & 3.08 & \textbf{0.72}\ & 0.60\ & 0.51\ & 0.55\ & 0.75\ &  96 \\

    Action Units (AU) & 3.02 & 0.67\ & 0.50\ & 0.35\ & 0.41\ & 0.67\ &  384 \\
    
    Eye-aspect ratio (EAR) & 3.37 & 0.64\ & 0.45\ & 0.41\ & 0.43\ & 0.63\ &  64 \\

    Inter-vector angle (IVA) & 3.57 & 0.59\ & 0.35\ & 0.28\ & 0.31\ & 0.55\ &  640 \\

    Top Significant Features (TSF) & 3.18 & 0.70\ & 0.55\ & 0.55\ & 0.55\ & 0.77\ & 27 \\

    Feature Selection (FS) & 3.08 & 0.69\ & \textbf{0.57}\ & \textbf{0.62}\ & \textbf{0.67}\ & \textbf{0.81}\ &  46 \\

    All features & \textbf{2.81} & 0.71\ & 0.59\ & 0.39\ & 0.47\ & 0.75\ &  1280 \\
    
    \bottomrule
 
\end{tabular}
\end{table}

Overall, the performance of the models varied, but the one using selected features showed the best results in predicting depression. From the AUROC plot (Fig. \ref{fig:roc_features_lopdo_lgbm}), we can observe even though model with HEA achieved better results in terms of accuracy of 72\%, the model itself is stable when combined with other features its yields much better results with more predictive performance stabilization.

\subsubsection{Minimum number of days needed to produce reliable detection}

\begin{figure}[h]
    \includegraphics[scale=0.35]{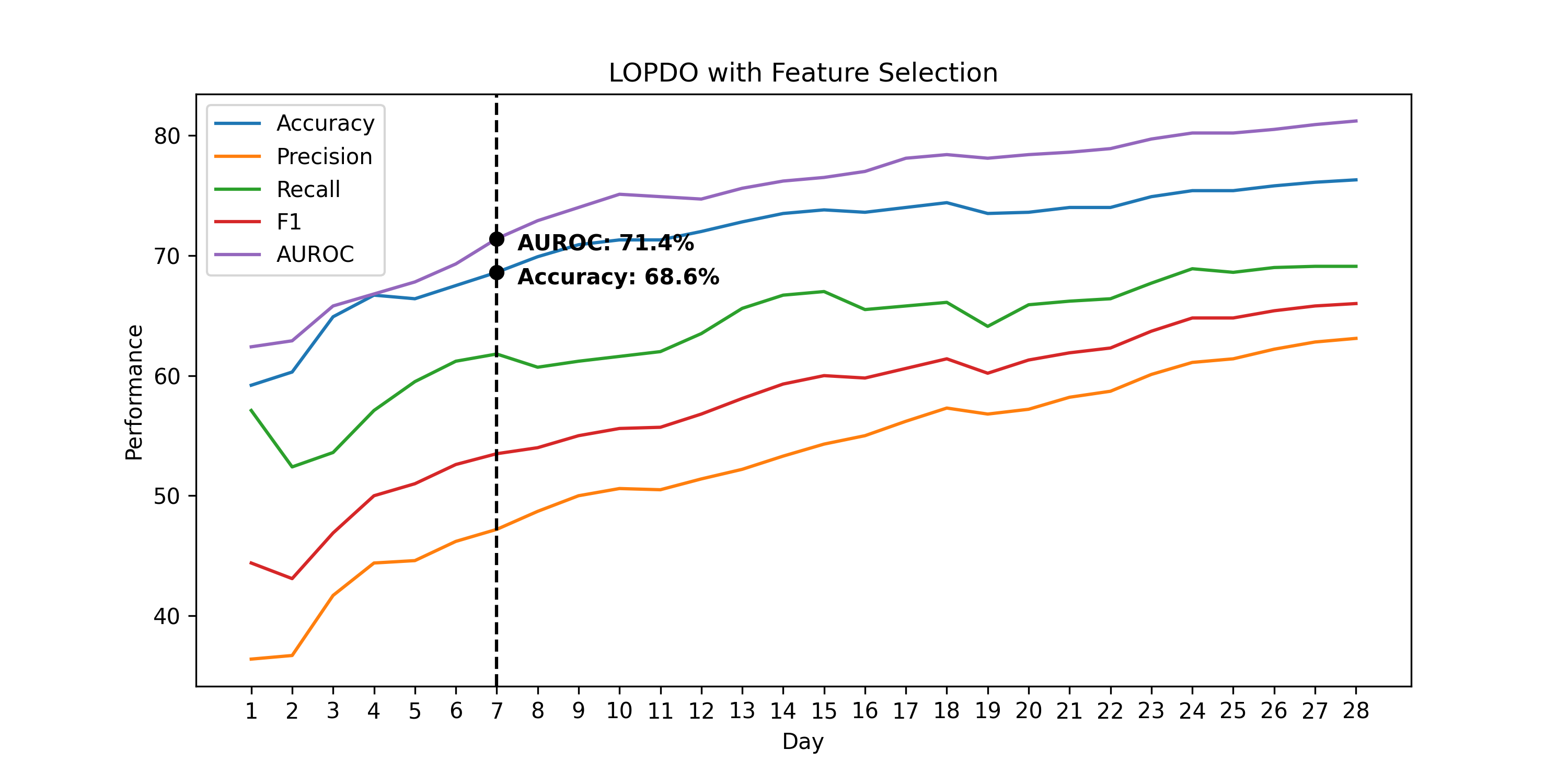}
    \caption{Minimum number of days needed to produce reliable detection}
    \label{fig:NumberofDaysPerformance}
\end{figure}

The AUROC is a metric used to evaluate the performance of a diagnostic test, with values ranging from 0.5 to 1. A value greater than 0.5 is necessary for the test to be meaningful, and an AUROC of 0.7 or above is generally considered acceptable. In the context of a depression detection model, the performance was better than random guessing on day 1, with an AUROC of greated than 0.5. On day 1, the model's performance improved to a fair level with an AUROC of 62.4\%. Remarkably, starting from day 7, the model achieved an acceptable performance with an AUROC of 71.4\%. This progression illustrates (Fig \ref{fig:NumberofDaysPerformance}) a significant enhancement in the model's ability to accurately detect depression over the weeks.

%% file: Section/6_Discussion.tex
\section{Lessons Learned: Discussing the Detection of Depressive Episodes on a Scale}
In this section, we highlight our work that contributes to new insights into the detection of depressive episodes in everyday settings by designing and developing an open-source affective mobile system. We found eye open state, head pose, smile, and action units (2, 6, 7, 12, 15, and 17) as key affective indicators in differentiating depressive and non-depressive episodes that have been validated in our field study. The combined features can be used to predict depression episodes. The universal model has an AUROC of 67\%, while the hybrid model has an AUROC of 81\%. Further improvement can achieved by collecting more data for subsequent weeks. These results serve as a bridge between controlled laboratory studies and real-world applications, demonstrating the feasibility of depression detection using an affective mobile sensing system.

\subsection{Comparison to Prior Work}
In the landscape of depression detection using mobile sensing, our approach significantly advances by leveraging in-the-wild data collection through everyday smartphone interactions. This method contrasts with many previous studies that primarily utilize lab-controlled \cite{cohn2009detecting, valstar2013avec, song2020spectral, kong2022automatic, casado2023depression} environments for data collection using camera modality, thus limiting the generalizability of their findings. For instance, our work complements the findings of Nepal et al. \cite{nepal2024moodcapture}, which also employs an in-the-wild approach but focuses on different feature sets, such as gaze and head pose, along with 2D and 3D facial landmarks. While their study achieves a balanced accuracy of 61\%, our method enhances the model's sensitivity to subtler indicators of depression through a detailed analysis of facial action units (AUs), eye-open states, and smile expressions. Our method achieves an AUROC of 81\% and Accuracy of 69\%, and maintains a consistent MAE across different validation strategies, highlighting the robustness of our findings and their potential for real-world application. This demonstrates the incremental utility of our approach, particularly in the seamless integration of mental health monitoring into daily technology use, thus contributing valuable insights into mobile health (mHealth) technologies for depression detection.

In comparison to mobile sensing, studies focused on depression detection, our approach capitalizes on direct behavioral markers accessible via a smartphone camera, distinguishing it from studies that rely on peripheral sensor data such as GPS or physical activity metrics. For example, Chikersal et al. \cite{chikersal2021detecting} and Opoku et al. \cite{opoku2022mood} employ a variety of sensors to infer depressive states indirectly through changes in mobility patterns and phone usage. While these studies achieve high accuracy and AUC scores, they may not capture the nuanced emotional states that facial behavior can indicate. Our method's utilization of detailed facial action units, eye-open states, and head gestures offers a more direct and potentially insightful measure of depressive episodes, evidenced by our comparable AUC of 81\%. Our model outperformed compared to a model that solely utilized sensor data from the AWARE platform \cite{ferreira2015aware}, which included Bluetooth (Accuracy=69.3, F1=0.64), Calls (Accuracy=68.5, F1=0.59), GPS (Accuracy=69.5, F1=0.62), and Steps Counter (Accuracy=63.6, F1=0.53), as detailed in research by Chikersal et al. \cite{chikersal2021detecting}. Our approach achieved an accuracy of 69\% and an F1 score of 0.67, surpassing individual sensor results. However, Chikersal et al. reported a higher F1 score of 0.78 when combining all sensors, indicating superior performance compared to our model in that specific setup. This specificity in detecting emotional expressions offers a critical enhancement over traditional mobile sensing methods, making our approach a valuable addition to the spectrum of technologies for monitoring mental health in everyday settings.

\subsection{Insights, Challenges and Opportunities in Predicting Depressive Episodes in the Naturalistic Environments}
Previous lab-controlled studies have delved into the extraction of facial behavior primitives from face images using affect sensing systems like OpenFace \cite{baltrusaitis2018openface}. These studies \cite{cohn2009detecting, valstar2013avec, song2020spectral, kong2022automatic, casado2023depression} have achieved impressive performance in detecting depression, due to their high data collection rate (e.g., processing video) and the controlled environment in which the data was collected. In contrast, deploying these systems in real-world scenarios using smartphones results in several challenges. On-device resource limitations often lead to a reduced frame rate (2.5Hz), and the unconstrained data collection settings -- affected by factors such as varied lighting conditions, phone orientation, ongoing activities, and whether the environment is indoors or outdoors -- can lead to poor quality of data collection. This, in turn, can impact the lack of samples that limits the development of predictive machine learning. Most recently, MoodCapture \cite{nepal2024moodcapture} was introduced that captures facial images in natural environments for depression detection. Their research demonstrated that using a random forest algorithm trained on facial landmarks, it's possible to identify depression and predict PHQ-8 scores among individuals. However, the utility of MoodCapture for developers intending to replicate such studies in different settings may be limited due to the lack of access to the mobile system, dataset, or machine learning framework used by the authors. Our research builds upon MoodCapture's work; we advance data collection by using a literature-based trigger mechanism that responds to user interactions like screen activity and app usage, enhancing user privacy by processing data on-device and discarding raw images post-analysis. Our study is novel in developing an open-source, privacy-aware mobile system that captures and processes facial data in near real-time, introducing significant improvements in privacy-awareness, data collection, and on-device processing.

The similarities and differences observed between lab-controlled and real-world predictive models can be attributed to the inherent nature of the environments in which they operate. Grounded in HCI/affective computing theories \cite{picard1999affective}, controlled environments allow for minimizing external variables \cite{campbell2015experimental}, ensuring that the subject's emotional state primarily influences the data collected. In such settings, the system can focus solely on the facial behavior primitives without interfering with external factors. However, in real-world scenarios, the myriad of uncontrollable variables, from lighting to personal activities, introduces noise into the data, which can mask or distort the true emotional indicators. From an affective computing perspective, human-computer interaction is dynamic and multifaceted in real-world settings. The system has to interpret the emotional state and account for the context in which the interaction is taking place. This context can significantly influence the emotional indicators, making them more complex to decipher. These insights are crucial for the design of an affective mobile framework for mental health. Recognizing the challenges of real-world data collection, future frameworks should incorporate adaptive algorithms that can adjust to varying conditions, ensuring consistent and accurate predictions. Additionally, leveraging context-aware computing can help the system contextualize the data, distinguishing between genuine emotional indicators and those influenced by external factors. This approach would lead to a more robust and reliable affective mobile sensing system, enhancing its potential in mental health applications.

Our method demonstrates good performance in depression detection compared to previous work by Opoku et al. \cite{opoku2022mood}, which used a similar learning scheme for a hybrid model. However, it's important to note that this comparison may not be entirely direct. Additionally, our universal model shows fair performance compared to prior work by Chikersal et al. \cite{chikersal2021detecting} in passive sensing using a similar learning scheme. The referenced study incorporates a comprehensive set of features, including sleep patterns, physical activity, phone usage, GPS location, and daily mood ratings, to model behavioral and social signals. However, as extensive research suggests, depression is a multifaceted disorder that affects various aspects of an individual's behavior, social interactions, and physiology. The existing mobile sensing-based approaches \cite{opoku2022mood, chikersal2021detecting, farhan2016behavior} have limitations in capturing affective signals \cite{abdullah2018sensing} manifested through involuntary facial muscle and head gestures, which have been established as crucial indicators of depression. Although wearable sensing-based physiological solutions have been proposed, their high deployment costs including extra cost for purchasing and wearing devices present a significant challenge and lack of affect/emotional signals. Previous research in affective computing conducted within controlled laboratory settings \cite{cohn2009detecting, valstar2013avec, song2020spectral, kong2022automatic, casado2023depression} has demonstrated significant promise in capturing emotional signals in individuals with depression. However, these studies are constrained in their applicability to real-world deployments due to computational cost, cost of devices, and user efforts to wear extra devices, particularly when continuous monitoring is essential for delivering timely interventions based on signals captured from users. The resolution of these issues remains unexplored. Therefore, our study aims to bridge these gaps by proposing, collecting, and evaluating the feasibility of deploying data using our novel affective mobile computing system in a real-world, naturalistic setting.

\subsection{Privacy and Ethical Considerations to Enhance Feasibility in Real-World Settings}
Because a camera temporarily captures a user's face on their smartphone for up to 10 seconds before deletion, users may have concerns due to the human perception of 'using a camera,' even though the system does not record any videos. To balance privacy considerations and data quality for modeling, researchers in the HCI community should consider designing user nudging \cite{balebako2014improving, felt2012ve}. These nudging, as suggested by researchers in privacy-preserving systems \cite{denning2014situ}, could allow users to push/pull status about system behavior when running in the background using content that includes appropriate information about data collection status in private contexts where they may feel uncomfortable, even when no images are collected/stored. 

While our system automatically removes images after near-real-time feature extraction, the concept of passive sensing \cite{denning2014situ} that underpins our affective mobile sensing framework does not necessarily require a user interface to reduce users' burden; instead, it runs in the background. This is in contrast to active sensing, typically used for manual tasks. Nevertheless, it is essential to empower users with the ability to enable or disable specific functionalities whenever they perceive potential risks, rather than requiring them to exit the study. While there is a trade-off between privacy and the quantity of data collected, it is crucial for HCI researchers to collaborate in designing mental health tracking systems that enhance user engagement \cite{o2008user}, encourage self-reflection \cite{li2010stage}, motivation \cite{consolvo2006design}, and trust \cite{kelley2009nutrition}, as has been well-established in the field of personal informatics. For example, our individual facial behavior and head pose features can provide users with daily happiness percentage scores using facial expression algorithms. This can serve as one of the motivational strategies that could encourage them to reflect on their mental condition and enable them to maintain good mental health practice, potentially contributing valuable data for long-term healthcare research in the field of mental health. 

To balance privacy considerations and data quality during system design, researchers should consider how data could be collected, used, and stored, as well as what implications this could have for the privacy of the users. While there may be discussion over how data processing can dictate/drive the level of invasiveness of the application when the users are given an option of choosing which type of processing or filters (presenting blurry face images for facial feature extraction) \cite{denning2014situ} they would allow the developers to carry out. At the same time, this necessary allowing the user to do so may impact the number of signals coming from the user's facial data for accurate behavior modeling.

We highlight the high monetary costs associated with wearable-based physiological markers and the lack of rich emotional data. While we agree with the observation regarding the financial aspect, we would want to adequately address the potential privacy costs of camera-based sensing, which collects user data opportunistically throughout the day. The privacy cost in this context refers to the potential risk of unauthorized access or misuse of personal and sensitive visual data captured by the cameras. A method to reduce this impact could involve discarding the user images immediately after computing the sensing values, thereby minimizing the storage of potentially sensitive information or performing the data processing in memory. Further, we would reduce the time for automatic feature extraction (currently 10 seconds per image). While we also provided notification to users that FacePsy is collecting data in the background, more comprehensive and clear justification that considers monetary and privacy costs and strategies to mitigate potential concerns. Facial data of a person contains various characteristics of the face such as shape (face landmarks, smile probability), eye shape (eye-aspect ratio), muscle movements (Action Units), and face orientation (head Euler angles). While these characteristics describe the state of the face at any given moment but don't reveal a person's identity.  It is important to ensure that any facial data gathered is encrypted and securely stored, as well as ensure that a user's identity is not revealed in any way. A utilitarian approach from the normative ethics point of view \cite{ruotsalainen2020health} to privacy-maintaining interactions with such data is to balance the benefits that can be gained from facial recognition technology with the potential privacy risks. This approach involves considering the trade-offs between the benefits of using facial recognition technology and the potential risks to privacy. This approach requires making decisions that prioritize the greater good, such as deciding to collect only the minimum amount of data necessary, processing the data in a way that does not reveal user identity, and encrypting and securely storing the data. Additionally, this approach may include taking steps to ensure that facial recognition technology is used responsibly and ethically, such as providing users with information on how their data is being collected and used and allowing them to opt out of the system if they choose to.

\subsubsection{User Feedback}

We collected feedback (questionnaire available in Appendix \ref{A_Study_Feedback})  from participants at the end of the study, providing valuable insights into their experiences and perceptions regarding the FacePsy app. Here, we explore the initial reactions, adjustments to the data collection processes, and the evolving acceptance of privacy measures throughout the study. This feedback is instrumental in understanding the real-world implications of deploying such technology and informs potential enhancements to improve user experience and trust. Below are detailed reflections across five key areas, supported by direct quotes from the participants, illustrating the nuanced reactions and suggestions for future development.

\begin{itemize}
     \item \textbf{Initial Perceptions and Consent}: Participants initially had mixed feelings about facial data collection. For instance, P10 stated, \textit{"I was a little skeptical my apps might hung due to this new functionality."} Despite initial hesitations, the consent process was generally found to be reassuring. P8 shared, \textit{"Yes, everything was properly addressed and I was well informed,"} reflecting a sentiment that the privacy and data usage concerns were adequately managed. Some participants (P23, P30) showed an initial discomfort with the app collecting their sensitive facial data regarding privacy. This discomfort may arise due to the introduction of new technology which they haven't used previously. While discomfort was mentioned by participants, they showed a shift in their comfort level using FacePsy over a period of use. 

    \item \textbf{Experience with Data Collection Triggers}: The experience of the app activating on phone unlocking or opening trigger apps varied. While some participants adjusted over time, others remained concerned. P35 commented, \textit{"It definitely became more acceptable over time. Even though I wasn't particularly concerned about data collection, I got more used to it after the first few days. "} showing a quick adaptation, while P24 noted, \textit{"It was a bit of an adjustment but became normal with time"}, showing slow adoption of apps trigger data collection in their daily smartphone use. Whereas P8 noted, \textit{"It was a slight concern at the start but I got used to it eventually."}

    \item \textbf{Impact on Daily Use and Privacy}: Changes in usage habits due to the app’s data collection were minimal. The feature that automatically discarded images after 20 seconds was positively received. For instance, P24 appreciated this feature, saying, \textit{"Increased comfort level"}, P23 mentioned \textit{"A bit of comfort, but I am also aware that social media companies convert the images into metadata. where the original image is no longer needed"}. This shows participants were educated about such data collection methods. While P35 mentioned, \textit{"It may have subconsciously led me to use my phone less frequently at beginning. But I got used to over time and got back to my normal phone usage habits. It didn't make me avoid using any specific apps."}

    \item \textbf{Understanding and Trust on-device Feature Extraction}: Understanding of how facial features were extracted varied, with some participants expressing a desire for more information. Trust was generally high for those who felt adequately informed. For example, P30 affirmed, \textit{"I trust it since it was made clear by the researchers that this was the case. I think as far as trust at the data collection, this is a great approach. As long as the image data is not leaving my phone, I do not have any issues with it."}. P35 showed increased comfort level with using their phone over time, noting \textit{"This was the main reason why I became more and more comfortable using my phone after the initial period. I think this was the key feature that made me stick with the study until the end."}. P23 advocated for a mechanism to create a private repository, noting \textit{"Create a private repository and also alert when the data is accessed with providing the reason and by whom it was accessed"} -- regarding the ownership and control of their personal data.

    \item \textbf{Long-term Acceptance}: Perceptions of the app improved over time for many participants. Regarding suggested improvements, P8 recommended, \textit{"I would make the activation not so random but in timed intervals"} calling for enhanced user control over the data collection processes. A similar sentiment was echoed by P35, \textit{"Giving users personalization options for which apps to exclude for data collection would be beneficial. Some users may prefer excluding certain apps from being monitored, and I think having this option would increase user acceptance."}
\end{itemize}

\subsection{Potential to Integrate Affective and Cognitive Inferences From Facial Behavior Markers with Conversational Artificial Intelligence (CAI)}
As previous research has confirm that understanding facial behavior primitives can enhance long-term solutions for managing depression and provide insights into emotional communication \cite{hammal2014intra}, aggression and negative affect recognition \cite{fitrianie2023head}, psychological distress \cite{stratou2017multisense}, and automatic thoughts perception during cognitive behavioral therapy (CBT) \cite{shidara2022automatic, shidara2020analysis}, inferences drawn from facial behavior primitives in our study could enable more effective affective interactions \cite{conati2005affective, islam2023revolutionizing} with a virtual CBT agent \cite{shidara2022automatic, shidara2020analysis}, depending on an individual’s emotional and cognitive state. Such an approach aims to create more affect-sensitive multimodal human–computer interactions \cite{pantic2003toward}, enhancing the human-like qualities, effectiveness, and efficiency of virtual agents. Real-time interpretation of complex mental states from facial expressions and head gestures, indicating concentration, fatigue, disagreement, interest, contemplation, uncertainty, and more \cite{el2005real}, could provide contextual affect data to virtual agents, facilitating more fluid interactions. This aspect is often lacking in text-based and avatar-based therapy agents.

In addition to its potential in depression detection, facial behavior primitives could find application in real-time intoxication detection, as demonstrated in previous work on drunk face detection using an offline dataset \cite{mehta2019dif}. By incorporating such capabilities, the FacePsy system could broaden its utility and use cases for real-world applications. Moreover, the system's applicability could extend to the detection of other forms of intoxication, such as marijuana intoxication \cite{bae2021mobile, chung2020mobile}, where observable facial muscle and head movements are indicative of psychomotor retardation and agitation. Recent developments, like the creation of an augmented reality feedback system for facial paralysis in a mobile setting \cite{barrios2021farapy}, hint at the potential utility of the FacePsy system in similar deployments. Beyond health-related applications, the FacePsy system's usefulness could be explored in monitoring student engagement in smartphone-based education or virtual classes \cite{islam2023microflow} where instructors can adjust lecture materials based on students' behavioral feedback. Indicators such as concentration, interest, or uncertainty are crucial for enhancing classroom engagement and delivering high-quality education. The FacePsy system represents a significant step in the realm of mobile affect sensing in human-computer interaction, with far-reaching implications for mental health with virtual therapy sessions, substance abuse detection, and educational engagement. 

%% file: Section/7_Limitation.tex
\section{Limitation and Future Work}
Even though we were able to get valuable insights about modeling depression and the proposed subset of facial and physiological signals, there are still improvements to be made for this system to be applicable in clinical settings. Although we successfully built a population model to detect depression, however, there might be individual patterns that our population model cannot capture, and thus may limit the generalizability of our model. In our future work, we will collect a larger dataset per participant and investigate the use of more personalized individual models. While we only register 11 categories of app use there could be more categories of app use that could work as the best avenue for data collection, further research should examine if such categories exist. The current limitation of our app, FacePsy, lies in its inadvertent triggering of data collection during intra-app navigation, such as moving between pages within the same app, leading to multiple data captures in a single session. In future work, we plan to refine the app's architecture to discern and limit data collection to significant user interactions, thereby enhancing the efficiency and relevance of the data collection process. Furthermore, in our future work, we want to integrate the pupillary response measurement module \cite{islam2024pupilsense} in our processing pipeline for in-app measurement of pupillary response by using Android Native libraries. In addition, we aim to enrich FacePsy by integrating it with systems like AWARE and Fitbit, combining rich emotional signals extracted from visual data with other data such as GPS, heart rate, and EDA. We also plan to add contextual layers, like categorizing apps that trigger data collection in model development which we collect as part of the FacePsy triggering mechanism, to deepen the understanding of facial behavior in context.

Our depression labeling strategy may raise concerns regarding the data's clarity and characteristics: Each session represents a unique data point with specific features. It's important to note that a single day could encompass multiple sessions. However, each session linked to a particular user will predict the different depressive episode labels span over two weeks observation period. This methodology presents two potential challenges. Firstly, the variability between sessions might make it challenging to identify a consistent pattern, which could affect the accuracy of predicting depressive episodes. Secondly, if sessions appear too homogenous, it could suggest that the model might be detecting implicit user characteristics rather than their actual risk of depression.

%% file: Section/9_Conclusion.tex
\section{Conclusion}
Depression is a major mental health disorder. As such, detecting depression can have significant impacts across several domains. Towards this goal, we propose an affective mobile system that allows us to collect facial behavior primitives from faces by opportunistically capturing user faces by observing the user interaction with their phone in a naturalistic setting. To this end, we build a depressive episode prediction model that achieves 81\% of AUROC. Our regression model that estimates PHQ-9 scores reached a moderate level of accuracy, exhibiting an MAE of 3.08. Based on the results from our cross-validation, we found our model produces reliable performance from several weeks of data to detect depressive episodes. We highlight key behavior primitives differentiating depressive and non-depressive episodes and use case scenarios regarding how the system could be applicable in detecting mental and neurological disorders for researchers and stakeholders. Lastly, we discuss privacy and ethical considerations in deploying such a system.

%% file: Section/10_Appendix.tex
\appendix

\section{Survey}

\subsection{PHQ-9}
Participants were asked: "Over the last two weeks, how often have you been bothered by the following problems?" This questionnaire is part of the standard assessment to gauge the severity of depressive symptoms. Below is the PHQ-9 questionnaire (Table \ref{tab:phq9}) used in the study:

\begin{table}[h]
\centering
\small
\caption{PHQ-9 Questionnaire Items}
\label{tab:phq9}
\begin{tabular}{>{\raggedright}p{0.5cm} >{\raggedright\arraybackslash}p{12cm}}
\toprule
\textbf{No.} & \textbf{Question} \\
\midrule
1 & Little interest or pleasure in doing things \\
2 & Feeling down, depressed, or hopeless \\
3 & Trouble falling or staying asleep, or sleeping too much \\
4 & Feeling tired or having little energy \\
5 & Poor appetite or overeating \\
6 & Feeling bad about yourself - or that you are a failure or have let yourself or your family down \\
7 & Trouble concentrating on things, such as reading the newspaper or watching television \\
8 & Moving or speaking so slowly that other people could have noticed. Or the opposite - being so fidgety or restless that you have been moving around a lot more than usual \\
9 & Thoughts that you would be better off dead, or of hurting yourself \\
\bottomrule
\end{tabular}
\end{table}

\subsection{Study Feedback} \label{A_Study_Feedback}
 Table \ref{tab:user_feedback} lists the user feedback questions administered at the end of the study to gauge participants' perceptions of the FacePsy app, focusing on aspects of consent, data collection triggers, impact on privacy, understanding of feature extraction, and long-term acceptance. The questions were designed to understand the participants' experiences throughout their interaction with the app and to gather suggestions for future improvements.

\begin{table}[h]
\centering
\small
\caption{User Feedback Questions}
\label{tab:user_feedback}
\begin{tabular}{>{\raggedright}p{0.5cm} >{\raggedright\arraybackslash}p{12cm}}
\toprule
\textbf{No.} & \textbf{Question} \\
\midrule
1 & When you first started using the FacePsy app, what were your initial thoughts about the facial data collection, especially when unlocking your phone or using specific apps? \\
2 & How were you informed about the data collection process, and did you feel that the consent process adequately addressed your concerns about privacy and data usage? \\
3 & Can you describe how you felt the first few times the app activated upon unlocking your phone or opening trigger apps? Did it become more acceptable over time, or did it remain a concern? \\
4 & Were there any particular trigger apps that made you more uncomfortable when the FacePsy app activated? How did this affect your usage of those apps? \\
5 & Did you notice any changes in your phone usage habits due to the app’s data collection methods? For example, did you use your phone less frequently or avoid certain apps? \\
6 & What are your thoughts on the app automatically discarding images after 20 seconds? Did this feature influence your comfort level with the ongoing data collection? \\
7 & How well do you understand the process of facial feature extraction by the app? Was there enough information provided about what data is extracted and how it is used? \\
8 & Do you trust that the facial data collected remains on your device and is not uploaded elsewhere? What could increase your trust in the system’s handling of your data? \\
9 & How has your perception of the FacePsy app and its data collection practices changed during the course of the study? \\
10 & What improvements or changes would you suggest for the app, especially regarding user control over data collection and privacy? \\
\bottomrule
\end{tabular}
\end{table}